\documentclass{aa}  

\usepackage{graphicx}
\usepackage{txfonts}
\usepackage{lipsum}
\usepackage{lscape}             
\usepackage{placeins}           
\usepackage{booktabs}                                
\usepackage{color}
\usepackage{natbib,twoopt}
\usepackage{xurl}
\usepackage[breaklinks]{hyperref}      
\bibpunct{(}{)}{;}{a}{}{,}             
\definecolor{cobalt}{rgb}{0.06, 0.2, 0.65}
\hypersetup{
  colorlinks,
  citecolor=cobalt,
  linkcolor=[rgb]{0.8, 0.2, 1.0},
  urlcolor=cobalt,
}
\makeatletter
  \newcommandtwoopt{\citeads}[3][][]{\href{http://adsabs.harvard.edu/abs/#3}%
    {\def\hyper@linkstart##1##2{}%
     \let\hyper@linkend\@empty\citealp[#1][#2]{#3}}}
  \newcommandtwoopt{\citepads}[3][][]{\href{http://adsabs.harvard.edu/abs/#3}%
    {\def\hyper@linkstart##1##2{}%
     \let\hyper@linkend\@empty\citep[#1][#2]{#3}}}
  \newcommandtwoopt{\citetads}[3][][]{\href{http://adsabs.harvard.edu/abs/#3}%
    {\def\hyper@linkstart##1##2{}%
     \let\hyper@linkend\@empty\citet[#1][#2]{#3}}}
  \newcommandtwoopt{\citeyearads}[3][][]%
    {\href{http://adsabs.harvard.edu/abs/#3}
    {\def\hyper@linkstart##1##2{}%
     \let\hyper@linkend\@empty\citeyear[#1][#2]{#3}}}
\makeatother

\newcommand{\Msun}{M$_{\odot}$}

\newcommand{\gppr}{\stackrel{>}{\scriptstyle \sim}}
\newcommand{\gappr}{\raisebox{-0.4ex}{$\gppr$}}

\usepackage{makecell}
\usepackage{siunitx}

\newcommand{\errnum}[2]{
\begin{tabular}[t]{@{}r@{}}
#1\\
$\pm$ #2
\end{tabular}
}

\begin{document}

   \title{Observational Constraints on Substellar Companions to High-Mass WDs with VLT/ERIS}
\titlerunning{Incidence of substellar companions to high-mass WDs}

%
   \author{
   Felipe Lagos-Vilches\inst{1}\fnmsep\thanks{Corresponding author: felipe.lagosv@usm.cl}
   \and Matthias R. Schreiber\inst{1}
   \and Alex Mauch-Soriano\inst{1,2}
   \and Wolfgang Brandner\inst{3}
   \and Hans Zinnecker\inst{4}
   \and Nicola Gentile Fusillo\inst{5,6}
    }

   \institute{
   Departamento de F\'isica, Universidad T\'ecnica Federico Santa Mar\'ia, Av. España 1680, Valpara\'iso, Chile
   \and 
   Instituto de F\'isica, Pontificia Universidad Cat\'olica de Valpara\'iso, Casilla 4059, Valpara\'iso, Chile
   \and
   Max Planck Institute for Astronomy, Königstuhl 17, 69117 Heidelberg, Germany
   \and
   Universidad Autónoma de Chile, Av. Pedro de Valdivia 425, Providencia, Santiago de Chile, Chile
   \and
   Department of Physics, Universitá degli Studi di Trieste, Trieste, 34127, Italy
   \and
   INAF-Osservatorio Astronomico di Trieste,Via G.B. Tiepolo 11, I-34143 Trieste, Italy
   }

   \date{Received September 30, 20XX}

 
  \abstract
   {The occurrence rate of substellar companions around intermediate-mass stars ($2\,\mathrm{M_{\odot}}\lesssim M \lesssim 8\,\mathrm{M_{\odot}}$) remains poorly constrained, partly because of the observational challenges associated with their high luminosities. White dwarfs (WDs) descended from these stars provide an alternative opportunity to investigate this population through direct imaging.}
   {We aim to constrain the occurrence rate of substellar companions around the remnants of intermediate-mass stars, providing complementary constraints on the companion population around their progenitors.} 
   {We conducted a high-contrast imaging survey of 19 WDs with masses between $0.68-1.22\,M_\odot$ and total ages of approximately $0.2$--$3$\,Gyr, using the NIX camera of VLT/ERIS in the $H$ band. We derived companion mass detection limits using ATMO evolutionary models and calculated the survey sensitivity. We also obtained second-epoch observations of selected candidates to assess their physical association with the WDs.}
   {We find no confirmed substellar companions in our sample. Our survey achieves detection probabilities above $50\%$ for companions more massive than $\sim12\,\mathrm{M_{Jup}}$ at semi-major axes of approximately $20$--$250$\,au, and yields a $95\%$ confidence upper limit on their occurrence rate of $\sim16\%$ at $50$--$100$\,au. By statistically accounting for potential merger products in our WD sample, the $95\%$ upper limit for WDs descended from single-star progenitors increases to $\sim26\%$.}
   {Our results provide new constraints on the occurrence of substellar companions around the remnants of intermediate-mass stars. The inferred occurrence rate is broadly consistent with previous observational constraints, although differences in sensitivity and orbital-separation coverage limit direct comparisons. Our non-detections may reflect a low intrinsic companion rate around the progenitors. However, formation mechanisms and subsequent orbital evolution may also play a role. While core accretion generally favours formation at closer separations, gravitational instability is expected to operate preferentially at wider separations. These companions may then migrate further outward during post-main-sequence mass loss, potentially placing them beyond the region probed by our observations. }


   \maketitle

\section{Introduction}

Understanding the occurrence rate of substellar companions as a function of host-star mass and orbital separation is key to constraining the formation of giant planets and brown dwarfs. While thousands of exoplanets have been discovered around main-sequence stars with masses comparable to or lower than that of the Sun, the occurrence rate and mass distribution of planets and brown dwarfs around intermediate-mass stars ($\sim2$--$8\,M_\odot$) remain poorly constrained, largely because of observational challenges. A- and B-type stars exhibit few, broad absorption lines owing to their high temperatures and rapid rotation, making radial velocity surveys inefficient. Their high luminosities and large radii also make the detection of companions through direct imaging and transits challenging.

Radial velocity surveys of evolved intermediate-mass stars offer an alternative approach \citep{wolthoff2022}. For stars with masses up to $\sim2\,M_\odot$, the frequency of Jupiter-like planets with orbital periods of a few hundred days increases with stellar mass, although this trend declines for more massive giants, while the metallicity dependence observed around lower-mass stars persists. At wide orbital separations, direct imaging surveys of young A- and B-type stars have provided complementary constraints on the occurrence of giant planets and brown dwarfs \citep{nielsen2013,Delorme2024}. Despite these efforts, the occurrence and mass distribution of substellar companions to stars more massive than $\sim3\,M_\odot$ remain poorly constrained, limiting our ability to test formation models in this regime.

The formation of substellar companions around intermediate-mass stars is of particular interest because it probes how planet formation depends on the mass of the host star. Around solar-type and lower-mass stars, core accretion (CA) is widely considered the leading mechanism for giant planet formation, involving the growth of a solid core followed by the accretion of a gaseous envelope \citep{Pollack1996,Emsenhuber2021}. However, the efficiency of this process depends on the properties and lifetime of the protoplanetary disc. The shorter disc lifetimes associated with more massive stars, together with their stronger photoevaporative radiation fields, may inhibit giant planet formation through CA, particularly for stars more massive than approximately $3\,M_\odot$ \citep[see][and most recently Ronco et al. 2026, A\&A submitted]{KennedyKenyon2008, Ronco2024} . Under these conditions, gravitational instability (GI), whereby sufficiently massive and rapidly cooling discs fragment into gravitationally bound clumps, may become an important alternative pathway for producing giant planets and brown dwarf companions \citep{Boss1997,KratterLodato2016}. Nevertheless, the relative contributions of these mechanisms and the characteristic masses of the resulting companions remain uncertain. Recent theoretical studies predict a range of outcomes, from the formation of a few-Jupiter-mass objects to populations dominated by brown dwarfs following subsequent accretion and dynamical evolution \citep{Ni2025,Schib2025}. Observational constraints on the mass distribution of substellar companions to intermediate-mass stars are therefore essential for testing these predictions.

A complementary approach to constraining the occurrence rate of substellar companions around intermediate-mass stars is to survey their stellar remnants: high-mass WDs ($M_\mathrm{WD}\gtrsim0.7\,M_\odot$). Their small sizes and low luminosities relative to their main-sequence progenitors facilitate the direct imaging of faint companions. Moreover, planets on orbits with semi-major axes of several au or more can undergo orbital expansion as their host stars lose mass during post-main-sequence evolution. For sufficiently wide orbits, this evolution is approximately adiabatic, and the final semi-major axis scales approximately with the ratio of the initial stellar mass to the final WD mass \citep{Villaver_2009,Nordhaus-13,Mauch-Soriano_2026}. Consequently, companions that initially orbited their host stars at separations of several au may end up at substantially wider orbits around the resulting WDs, making them more accessible to direct imaging. Finally, the thermal evolution of companions to intermediate-mass stars may also favour their detection at later evolutionary stages. For instance, higher equilibrium temperatures experienced during the main-sequence phase may delay their cooling, potentially enhancing their detectability compared with isolated substellar objects.

Despite these advantages, searches for substellar companions to WDs have yielded relatively few detections. The known population spans a wide range of orbital configurations, from close-in planets that may have survived or been delivered through post-main-sequence interactions \citep{vanderburgetal20-1,munoz+petrovic20-1,lagosetal21-1,maldonadoetal21-1,Chamandy-21,Stephan-21} to objects on wider orbits whose evolutionary histories are less certain \citep{luhman_2012,Gaensicke2019,blackmanetal21-1}. These systems demonstrate that substellar companions can survive stellar evolution and be present around stellar remnants, but they do not yet provide a statistically complete picture of their occurrence rate.

Several searches for substellar companions to WDs have been conducted using direct imaging, infrared excess measurements, and transit observations \citep{Debes2005,Farihi2005,Farihi2008,Hogan2009,Steele2011,Girven2011,vanSluijs_VanEylen2018,Brandner2021}. More recently, ongoing James Webb Space Telescope programs, including the MEAD survey \citep{Albert2026}, are further expanding the observational constraints on substellar companions to WDs, including cold giant planets.

Here, we present a systematic high-angular-resolution imaging survey for substellar companions around 19 young-to-intermediate-age ($\approx0.2$--$3$ Gyr), high-mass WDs using the Near-Infrared Camera System (NIX) of the Enhanced Resolution Imager and Spectrograph \citep[ERIS;][]{Davies2023} at the Very Large Telescope (VLT). The survey primarily, but not exclusively, targets WDs descended from stars with initial masses above $\sim3\,M_\odot$. Our observations are particularly sensitive to giant planets and brown dwarfs at wide orbital separations, providing observational constraints on the occurrence rate of substellar companions that can be used to test predictions from formation models, including GI. We identify several companion candidates; however, follow-up observations and statistical considerations indicate that none can currently be confirmed as physically associated with their host WDs. We therefore derive upper limits on the occurrence rate of substellar companions as a function of companion mass and orbital separation. These results provide further observational constraints on the demographics of substellar companions to intermediate-mass stars.

\section{Target selection}

Using the Montreal WD database \citep{Dufour2017}, we compiled an initial sample of WDs with masses above $\approx0.65$\,\Msun (corresponding to progenitor masses above $\approx 2$\,\Msun ), 
astrometric distances within $100$\,pc, and no previous diffraction-limited, high-contrast observations with the VLT. We also excluded WDs with known stellar or substellar companions, as well as those showing evidence of magnetism that may indicate a merger origin \citep[e.g., ][]{Tout2008, Briggs2015}. 

We then used the Python package \texttt{wdwarfdate} \citep{Kiman2022} to estimate the total age $\tau_\mathrm{tot}$ of each WD in our initial sample. The total age is calculated as the sum of the WD cooling age $\tau_\mathrm{cool}$ and the main-sequence lifetime of its progenitor $\tau_\mathrm{MS}$.
\texttt{wdwarfdate} requires estimates of the WD effective temperature $T_\mathrm{eff}$ and surface gravity $\log g$, from which it derives the WD mass,  $M_\mathrm{WD}$, and cooling age using cooling-track models. For this purpose, we adopted the $T_\mathrm{eff}$ and $\log g$ measurements from \citet{Vincent_2024}. The progenitor mass, $M_\mathrm{prog}$, is then inferred through an initial-to-final mass relation (IFMR), followed by its main-sequence lifetime from the MIST isochrones \citep{Choi_2016, Dotter_2016}.

To account for uncertainties associated with the choice of cooling models and IFMR, we considered multiple combinations of these prescriptions. In addition to the Montreal WD cooling tracks \citep{Bedard2020}, we incorporated the cooling models from La Plata WD group \citep{Camisassa_2016, Camisassa_2017, Camisassa_2019}. We also adopted the empirical IFMRs of \citet{Cummings_2018} and \citet{Miller_2026}, both based on WDs in open clusters, as well as the empirical IFMR of \citet{Cunningham_2024}, derived from field WDs within 40 pc of the Sun. The combination of two sets of cooling tracks and three IFMRs yields six different estimates of $\tau_\mathrm{tot}$, $M_\mathrm{WD}$ and $M_\mathrm{prog}$ for each WD.

To avoid numerical artifacts associated with the interpolation of cooling models at uncertainties smaller than the effective grid resolution, we adopted a minimum uncertainty of 1\% in effective temperature and 0.03 dex in surface gravity. The final age assigned to each WD was taken as the median of the six total-age estimates. The associated dispersion was quantified by the range spanned by these estimates, with the upper and lower bounds defined as the differences between the median values and the maximum and minimum values obtained from the six model combinations, respectively.

We restricted the final sample to WDs with total ages $\leq3$\,Gyr. This age cut was motivated by the expected detectability of substellar companions: according to the ATMO evolutionary models \citep{Phillips_2020}, companions with masses $\gtrsim10\,\mathrm{M_{Jup}}$ around WDs within this age range are expected to remain sufficiently luminous to be detectable at angular separations $\gtrsim1.0''$, with a signal-to-noise ratio $\gtrsim7$. The details of the observing strategy and its sensitivity are presented in Section~\ref{sec:Observations and data reduction}.

\begin{figure}
    \centering
    \includegraphics[width=\columnwidth]{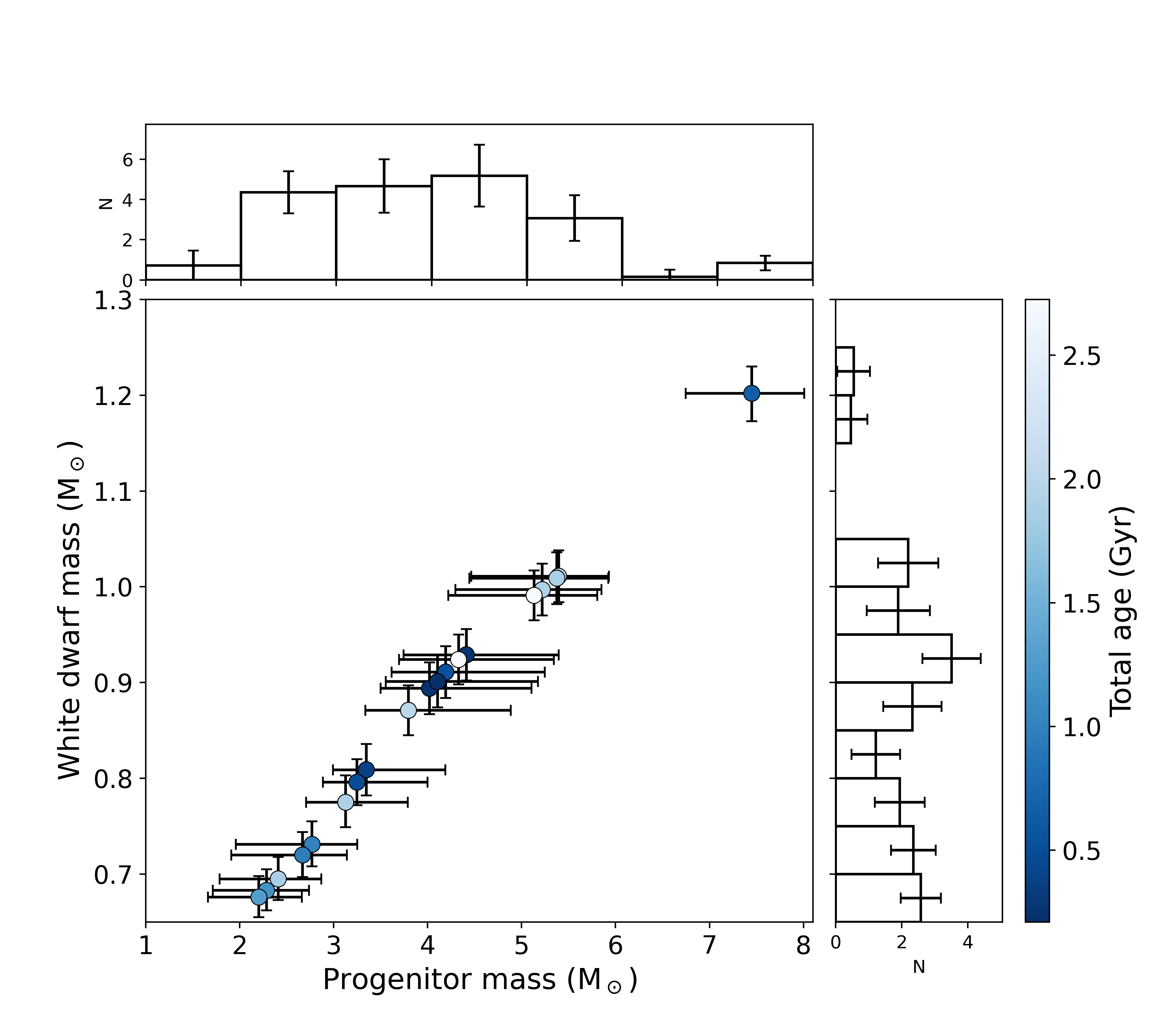}
    \caption{WD mass versus progenitor mass ($\overline{M}_\mathrm{WD}$ and $\overline{M}_\mathrm{prog}$; Table~\ref{tab:masses}) derived with \textsc{wdwarfdate}, combining three IFMRs and two WD cooling models. The inferred total system ages ($\overline{\tau}_\mathrm{tot}$; Table~\ref{tab:total_ages}), obtained by adding the WD cooling age to the progenitor main-sequence lifetime, range from approximately $0.2$ to $3$\,Gyr.}
    \label{fig:mass_dist}
\end{figure}

Figure~\ref{fig:mass_dist} and Tables~\ref{tab:total_ages} and \ref{tab:masses} show the total ages, WD masses, and progenitor masses of the final sample, consisting of 19 high-mass WDs.
The total ages range from $\approx0.2$ to $3$\,Gyr, while the WD masses span $\approx0.68$--$1.22$\,\Msun. The progenitor masses range from $\approx2$ to $8$\,\Msun, with most lying between $2$ and $6$\,\Msun.

\section{Observations and data reduction}
\label{sec:Observations and data reduction}
We obtained high-angular-resolution images of the 19 WDs using the ERIS/NIX camera during periods 111, 113 and 114. All targets were observed in the $H$ band (centred at $\lambda_{\rm c}=1.66\,\mu$m) using the small-field camera configuration, which provides a field of view (FoV) of $26.4''\times26.4''$ and a pixel scale of 13\,mas\,pix$^{-1}$. Observations were carried out in laser guide star adaptive optics (LGS-AO) mode, using the laser guide star to measure atmospheric distortions and the WD itself as the natural guide star for tip-tilt correction.

The observing strategy was designed to obtain a total integration time of 1728\,s per target while avoiding saturation of the WD. This was achieved using detector integration times of 8\,s and 18 integrations per exposure (DIT = 8\,s, NDIT = 18), combined with three dither positions, two instrumental position angles on the sky, and two detector offsets for PSF and background subtraction. The two position angles were $0^\circ$ and $20^\circ$ for period 111, and $0^\circ$ and $35^\circ$ for periods 113 and 114. This observing sequence yielded 12 data cubes per target, each containing 18 images.

The data cubes were first reduced using the \texttt{eris\_nix\_img} workflow available within the \texttt{EsoReflex} environment \citep{esoreflex_2013}, including dark subtraction, linearity correction, and flat-fielding. Bad-pixel correction was subsequently applied using the bad-pixel map produced by \texttt{eris\_nix\_img} as input to the \texttt{fix\_badpix} Python function from the VLT/SPHERE pipeline\footnote{\url{https://github.com/avigan/SPHERE/blob/master/sphere/utils/imutils.py}}. This routine replaces bad pixels by a distance-weighted average of the 12 nearest valid neighbouring pixels. At this stage, the post-processing of each image is performed on a $1050\times1050$ pixel patch ($13.65''\times13.65''$) centred on the WD to ensure a symmetric region around the target while avoiding the detector edges and regions affected by a high density of bad pixels.
Each patch was further corrected for outlier pixels (e.g. cosmic rays) using the \texttt{cube\_fix\_badpix\_clump} function from the Python package VIP\footnote{\url{https://github.com/vortex-exoplanet/VIP/blob/master/src/vip_hci/preproc/badpixremoval.py}} \citep{Gomez-Gonzalez2017, Christiaens_2023}, which flags a pixel as bad if its value deviates by more than 5 standard deviations from the median computed within a neighbouring region of 5-pixel radius, and replaces it with the local median.

\begin{figure}
    \centering
    \includegraphics[width=\columnwidth]{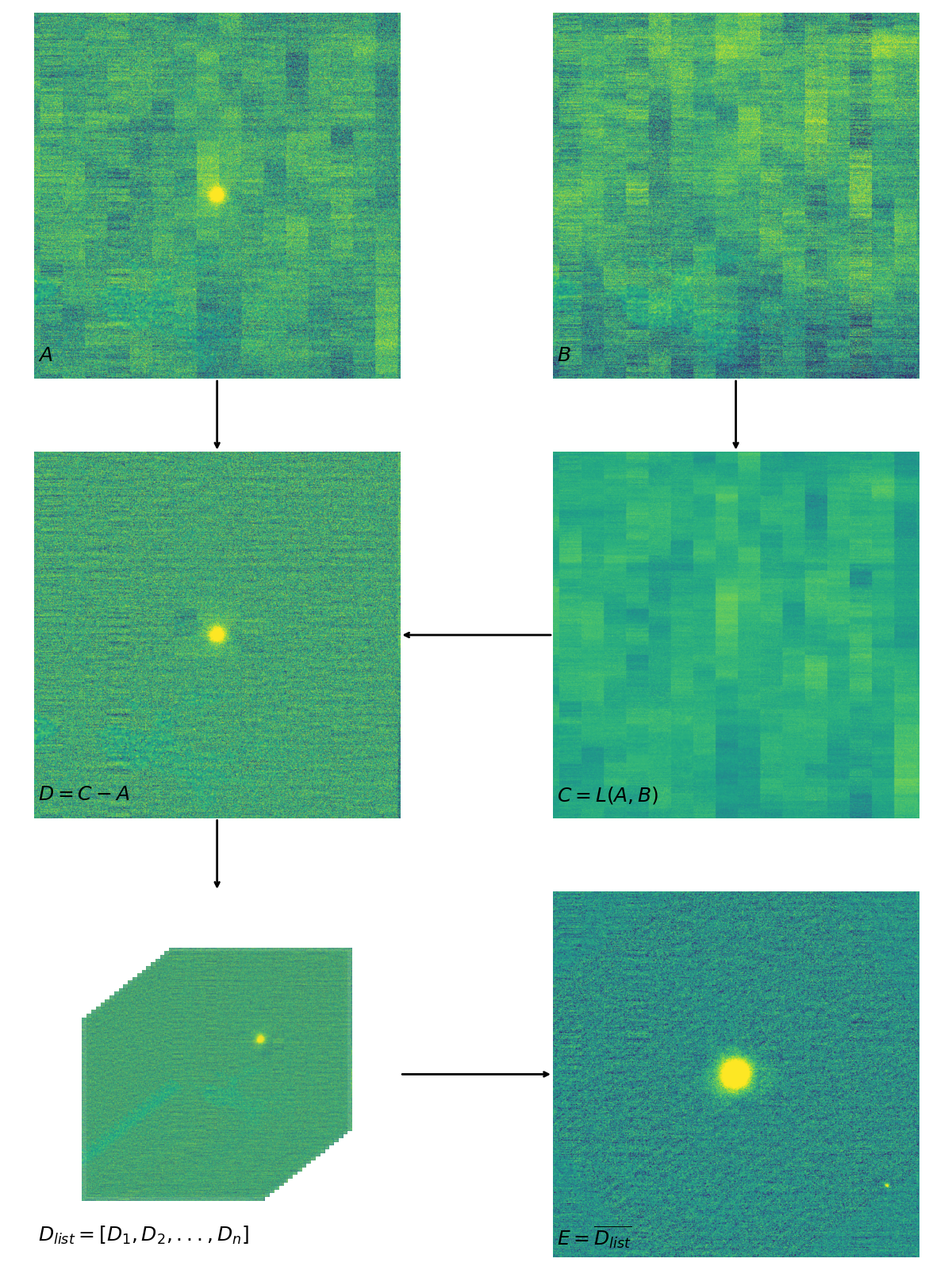}
    \caption{Flowchart illustrating the post-processing pipeline applied to the $1050\times1050$ pixel patches centred on each WD. (A) Output image from EsoReflex. (B) Master artifact model. (C) Linear fit of the master artifact model to the object patch. (D) Object patch after subtraction of the fitted artifact model. ($D_{\mathrm{list}}$) Patches are aligned, de-rotated to a common orientation of $0^\circ$, and stored in a data cube. (E) Patches with the best adaptive-optics performance are selected from the data cube and mean-combined to produce the final science image.}
    \label{fig:pipeline_steps}
\end{figure}

As shown in the top left panel of Fig. \ref{fig:pipeline_steps}, the resulting images (henceforth object images) still display instrumental artifacts, including vertical fringes caused by random offsets in the 64-pixel-wide readout channels, as well as features introduced by detector non-uniformities. We correct for these effects and perform the sky subtraction as follows. First, we remove the sky signal and random offsets by subtracting, from each fringe, its median pixel value. Any visible sources, including the WD, are masked prior to the median calculation. Second, for each median fringe-subtracted object image, we generate and subtract a model of the instrumental artifacts. To build this model, we first create a master-artifact frame by median-collapsing a set of median fringe-subtracted $1050\times1050$ pixel patches, centred at the detector coordinates corresponding to the WD position in the complementary offset positions. The artifact model for each image is then obtained by performing a linear fit of the master artifact frame to the corresponding median fringe-subtracted object image. This fitting is carried out independently within boxes defined by dividing each fringe into segments of 32 pixels in height (26 pixels for the last segment). For the boxes containing the WD, the fitted parameters are linearly interpolated from those within the same fringe. Additionally, any other previously detected sources are masked during the fitting process.

From the final set of post-processed object images, we select those with the best adaptive-optics performance using two metrics: the effective full width at half maximum ($\mathrm{FWHM_{eff}}$), defined as twice the radius enclosing $50\%$ of the WD point-spread-function (PSF) flux, and the encircled energy ratio ($\mathrm{EE_{ratio}}$), given by the flux PSF core relative to that in the halo. The core is defined within $1.5\times\mathrm{FWHM_{eff}}$, while the halo extends out to $6\times\mathrm{FWHM_{eff}}$. We retain frames with $\mathrm{FWHM_{eff}}$ below the 70th percentile and $\mathrm{EE_{ratio}}$ above the 70th percentile, selecting images with both sharper PSFs and high core-to-halo flux concentration.

The final set of images is aligned to the WD position in the first frame, de-rotated to a position angle of $0^{\circ}$, and then mean-combined. Figure \ref{fig:pipeline_steps} shows the main steps involved in the data post-processing and the resulting final image.

\section{WDs with Companion Candidates}

Among the observed WDs, we identified several companion candidates, most of which are expected to be unrelated background sources. To assess the likelihood of a chance alignment, we modelled the occurrence of background objects as a Poisson process and estimated the probability, $P_\mathrm{bkg}$, of finding at least one background source brighter than the candidate magnitude, $H_\mathrm{candidate}$, within the angular separation, $\Theta$, of each candidate.

If $\rho(>H_\mathrm{candidate})$ is the cumulative surface density of background sources brighter than $H_\mathrm{candidate}$, the probability of a chance alignment is \citep[e.g.,][]{Brandner2000}

\begin{equation}
P_\mathrm{bkg}(\Theta,H_\mathrm{candidate})=
1-\exp\left[-\pi\Theta^2\rho(>H_\mathrm{candidate})\right].
\end{equation}

The background population was assumed to consist of both Galactic stars and background galaxies. The stellar surface density, $\rho_\mathrm{stars}$, was estimated with the Besançon Galactic population synthesis model \citep{Robin_2004}, generating synthetic 2MASS catalogues over areas between 1 and 10 square degrees (depending on the stellar density required to minimize Poisson noise to $\lesssim1\%$), centred on the coordinates of each WD.

The contribution from background galaxies was estimated using the H-band galaxy number counts of \citet{Metcalfe_2006}, derived from deep extragalactic surveys. Unlike the stellar background, which was computed individually for each target, the same galaxy counts were adopted for all WDs, since they depend on limiting magnitude rather than sky position. Assuming conservatively that all background galaxies are unresolved in the ERIS/NIX images, these counts provide an upper limit to the expected surface density of extragalactic contaminants.

As an empirical reference for identifying promising companion candidates, we considered the WD LAWD\,83, which was observed with JWST/MIRI in April 2023 \citep{Mullally_2024}. By comparing the MIRI F560W images with our ERIS H-band observations, we confirmed that all three companion candidates, each with $P_\mathrm{bkg}>10\%$, are non-comoving background sources. Motivated by this result, we adopt $P_\mathrm{bkg}<10\%$ as a practical threshold for prioritizing candidates for follow-up observations. Across the full sample, only the innermost candidate around UCAC4\,460--008869 and the candidate around WD\,0136--201 satisfy this criterion, making them the most promising candidates for physical companionship.

\begin{figure}
    \centering
    \includegraphics[width=\columnwidth]{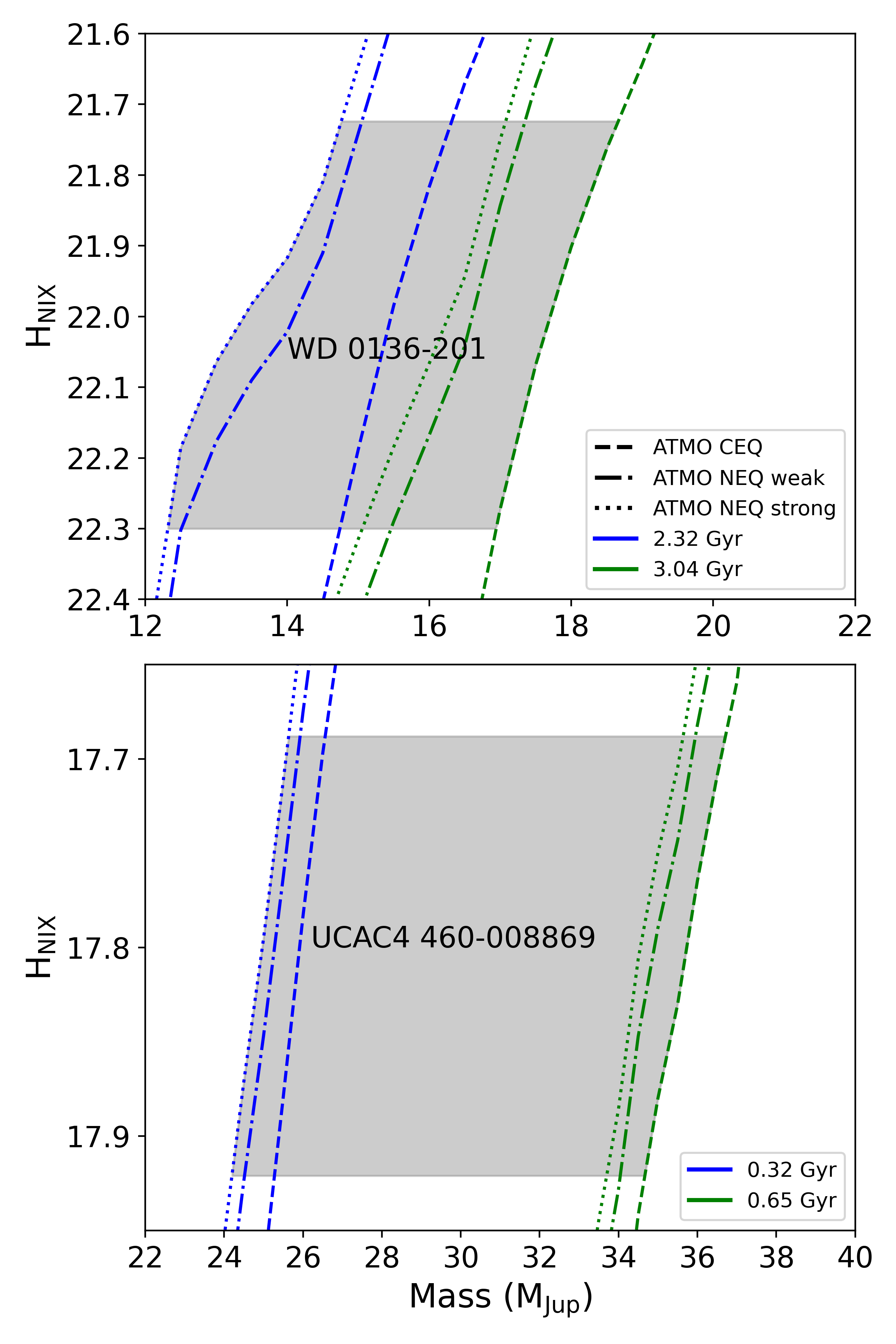}
    \caption{ATMO H-band magnitude–mass relations for the companion candidates around WD\,0024--556 and UCAC4\,460--008869, assuming system ages corresponding to the \(1\sigma\) lower (blue) and upper (green) limits of the total WD ages. The grey shaded regions indicate the range of companion masses allowed by the \(1\sigma\) uncertainties in both the H-band photometry and the system age.}
    \label{fig:TOIs}
\end{figure}

Assuming that both candidates are physically associated with their host WDs, we used their ERIS/NIX H-band photometry together with the ATMO evolutionary models to derive preliminary estimates of their masses and effective temperatures. For WD\,0136--201\,b we obtained a mass of $12$--$18.6\,M_\mathrm{Jup}$ (see top panel of Fig. \ref{fig:TOIs}) and effective temperature of $424$--$459$\,K. If confirmed, this object would represent the first known substellar companion to a massive WD (i.e. $M_\mathrm{WD}>0.9,M_\odot$) and only the second T/Y-type dwarf discovered around a WD, after WD\,0806--661\,B \citep{luhman_2012}. For UCAC4\,460--008869 b, we obtained a mass of $24-36.7\,M_\mathrm{Jup}$ (bottom panel of Fig. \ref{fig:TOIs}) and an effective temperature of $1179-1298$\,K. It is worth to mention that \cite{Xu_2020} reported a $\approx 3.5\sigma$ W1-W2 colour excess in the unWISE data for UCAC4\,460--008869, which could be associated with the companion candidate now resolved in our ERIS/NIX observations.

\section{Follow-up observations of companion candidates around high-proper-motion WDs}

The high proper motion of WD\,0136--201 ($\sim266\,\mathrm{mas\,yr^{-1}}$) makes it possible to assess the physical association of its companion candidate through a common proper-motion test. Follow-up observations with ERIS/NIX were carried out on 22 September 2026, providing a nearly-two-year baseline over which the WD is expected to have moved by approximately $0.53''$. We also observed the candidate around WD\,0024--556 on the same date. Despite its relatively high background probability ($P_\mathrm{bkg}\approx39\%$), we prioritized this candidate for follow-up because, if physically associated, it would have properties similar to those inferred for the candidate around WD\,0136--201, and the  high proper motion ($\sim542\,\mathrm{mas\,yr^{-1}}$) of WD\,0024--556.

\begin{figure}
    \centering
    \includegraphics[width=\columnwidth]{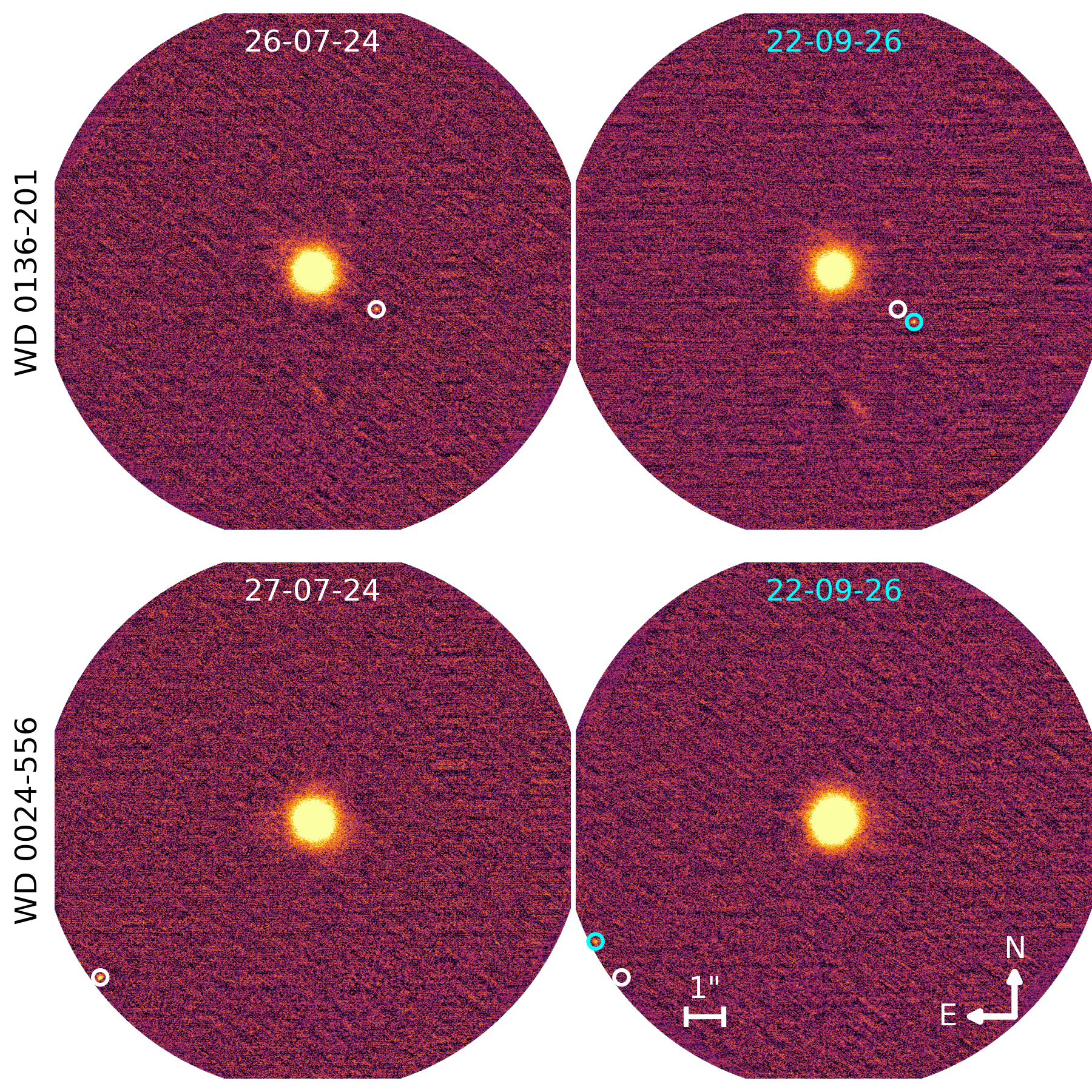}
    \caption{First- and second-epoch observations of WD\,0136--201 (top) and WD\,0024--556 (bottom). The positions of the companion candidates at the first and second epochs are marked with white and cyan circles, respectively.  If the candidates were physically associated with their host WDs, the circles would overlap in the second-epoch images. The clear positional offsets observed for both candidates indicate that they are background objects.}
    \label{fig:followup}
\end{figure}

Figure \ref{fig:followup} shows that the observed positions of both candidates are consistent with those expected for stationary background sources, given the proper motions of their respective host WDs. These results indicate that both candidates are likely unrelated background objects rather than physically bound companions.

\section{Constraint on the population of substellar companions to WDs}

To derive observational constraints on the population of substellar companions to massive WDs, we first determine contrast curves and convert them into limiting magnitudes. Using substellar evolutionary models and estimates of the host stars' ages, we then translate these magnitude limits into minimum detectable companion masses. Finally, we combine the resulting mass limits to construct detection probability maps which are then used to estimate limits on the the occurrence rate. Each of these steps is described in detail below.

\subsection{Contrast curves}
For each mean-combined WD image, we compute radial sensitivity limits following the methodology of \citet{Mawet_2014}. Briefly, at discrete separations of $r = 2n \times \mathrm{FWHM_{eff}}$ ($n=1,2,3,\ldots$), we measured the flux within a set of equally spaced circular apertures centered at the same radial distance from the WD. Each aperture had a radius of $r_\mathrm{aper}=\mathrm{FWHM_{eff}}$.

At each separation, the $5\sigma$ detection threshold was estimated as the mean flux of the apertures plus five times their standard deviation. The resulting threshold was corrected for small-sample statistics using the Student's $t$ correction described by \citet{Mawet_2014}. The corrected $5\sigma$ noise curve was then normalized by the WD flux, measured using an identical aperture, and converted into a contrast curve in magnitude units. As shown in the top panel of Fig. \ref{fig:contrast_curves},
the mean $5\sigma$ contrast stabilizes at approximately $8.3 \pm 1.0$ mag at separations $\gtrsim1.5''$ from the WD.

\subsection{Apparent magnitude detection limits}

\begin{figure}
    \centering
    \includegraphics[width=\columnwidth]{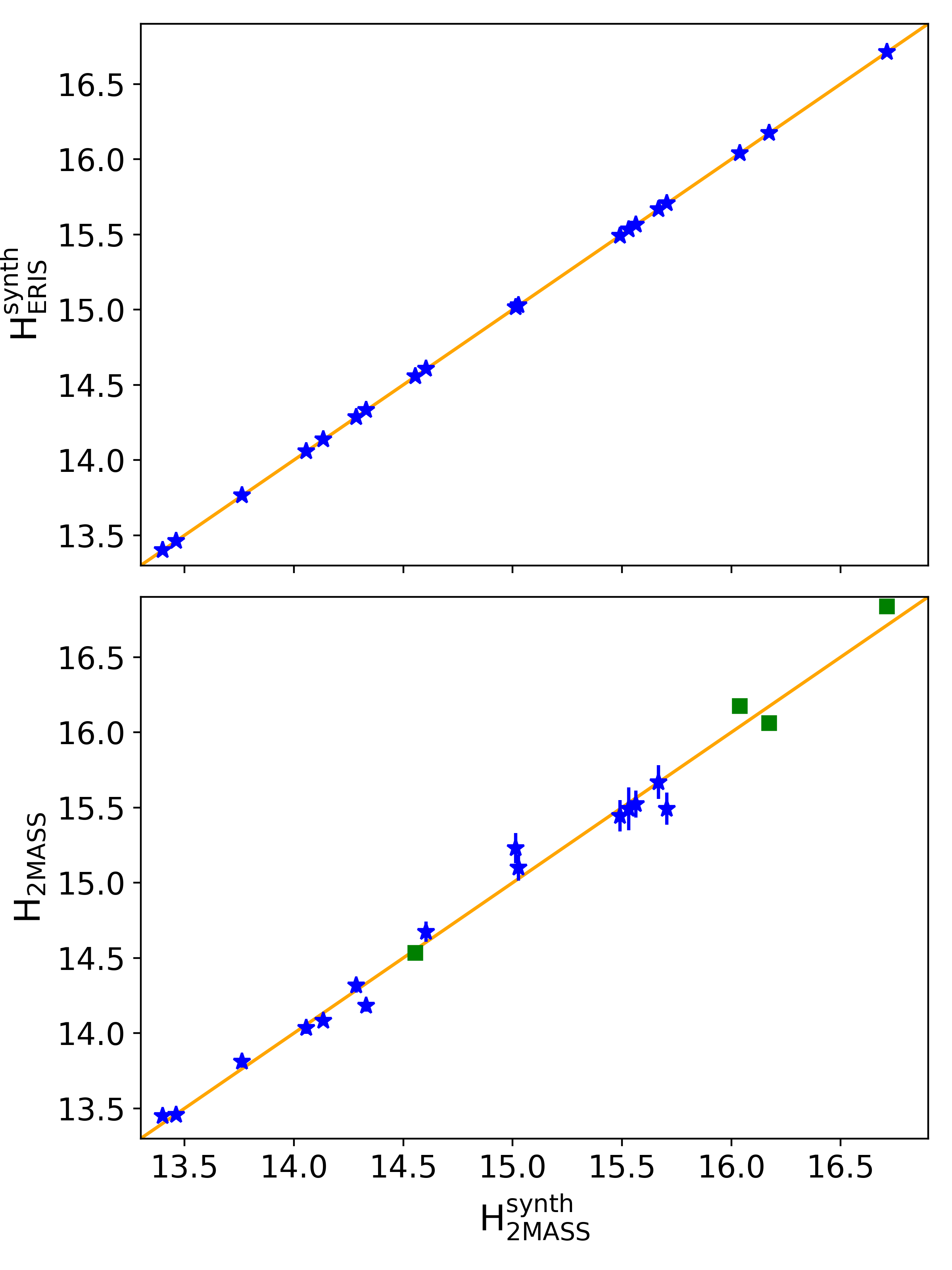}
    \caption{\textbf{Top}: Comparison between the ERIS/NIX and the synthetic 2MASS H-band photometry, showing a 1:1 relation (yellow solid line) that supports using the 2MASS photometry as a photometric reference (average discrepancy of $10^{-3}$\,mag). \textbf{Bottom}: Archival versus synthetic 2MASS H-band photometry for each WD in our sample. All values agree within $3\sigma$ of the observed uncertainties or $0.14$\,mag. Green squares mark WDs whose 2MASS H-band photometry was derived from SPHEREx data. }
    \label{fig:2mass_eris_Hmag_relation}
\end{figure}

\begin{figure}
    \centering
    \includegraphics[width=\columnwidth]{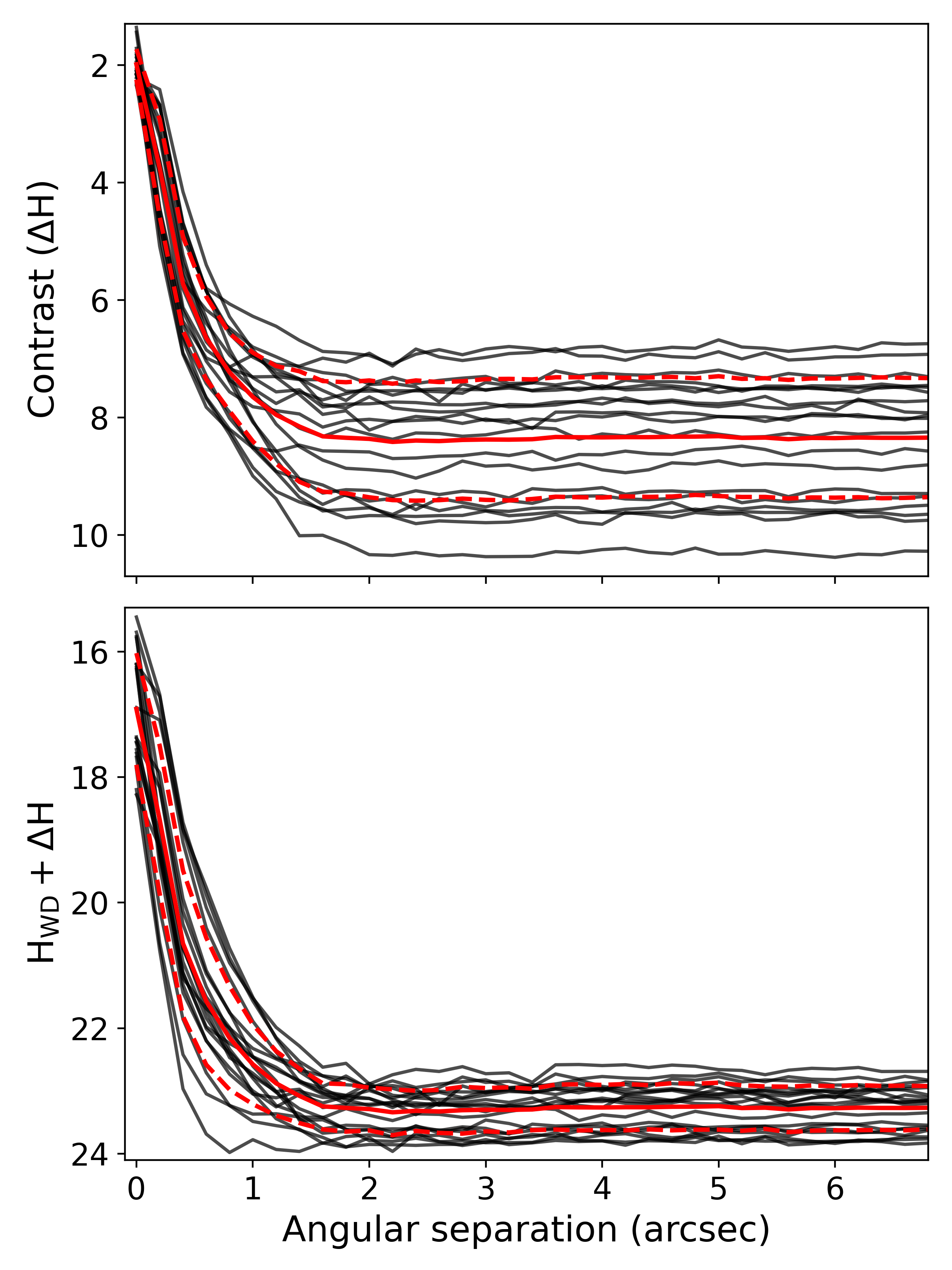}
    \caption{\textbf{Top}: $5\sigma$ contrast curves for the 19 WDs in our sample (black lines). The solid and dashed red lines represent the mean contrast curve and its $\pm1\sigma$ scatter, respectively. \textbf{Bottom}: $5\sigma$ contrast curves from the top panel converted to limiting H-band magnitudes using the 2MASS H-band magnitudes of the WDs. The survey reaches limiting magnitudes of $\approx23$ mag at angular separations $>1.5''$.}
    \label{fig:contrast_curves}
\end{figure}

The $5\sigma$ contrast curves were converted into H-band apparent-magnitude detection limits using the WD H-band magnitudes from the 2MASS All-Sky Catalog of Point Sources \citep{Cutri_2003} as a photometric reference. To assess the validity of using 2MASS H-band photometry as a proxy for the ERIS H-band filter, we employed the Python package \texttt{species} \citep{Stolker_2020} to compute synthetic photometry for both filter transmission curves across our WD sample. Synthetic magnitudes were calculated using the WD model spectra of \citet{Koester_2010} together with the $T_\mathrm{eff}$ and $\log g$ values listed in Table~\ref{tab:WD_sample}. The top panel of Fig.~\ref{fig:2mass_eris_Hmag_relation} shows that the magnitudes derived from the two filters are nearly identical, following an almost perfect one-to-one relation with negligible offsets. This agreement is expected given the broad-band nature of both filters and the relatively smooth spectral energy distributions of WDs across the H-band wavelength range.

We further tested how well the synthetic 2MASS H-band magnitudes reproduce the observed 2MASS photometry. We found a mean absolute discrepancy of $0.08 \pm 0.06$ mag, showing that the synthetic magnitudes reproduce well the observed values. 

Three WDs in our sample (GD\,1706, EC\,04426$-$3523, and GALEX\,J081835.6$-$183243) have poor-quality 2MASS H-band photometry. For these systems, we computed and adopted synthetic 2MASS H-band magnitudes using \texttt{species} and low-resolution near-infrared spectrophotometry obtained with the SPHEREx Spectrophotometry Tool \citep{Akeson_2025}. The spectrophotometric data were extracted from the SPHEREx Quick Release 2 Spectral Images \citep[QR2;][]{SPHEREx_QR2_2025} of the SPHEREx mission \citep{Bock_2026}. 

In addition, during the photometric analysis of RX\,J1847.9$-$2219, we found a significant discrepancy between its catalogued 2MASS H-band magnitude and the synthetic magnitude predicted from its WD model spectrum. We therefore estimated its H-band magnitude independently using the SPHEREx spectrophotometry, obtaining excellent agreement with the value derived from the model spectrum. Since RX\,J1847.9$-$2219 is located in a relatively crowded field (Fig. \ref{fig: gallery}), the original 2MASS photometry may be affected by contamination from nearby sources.

Bottom panel of Fig. \ref{fig:_mass_detection_limit} show the final $5\sigma$ H-band detection limits for the WD sample, which asymptotically reaches $H\sim23.3 \pm 0.4$ mag beyond $1.5''$. This magnitude limit is in good agreement with the ERIS Exposure Time Calculator estimate, which indicates that point sources with $H \approx 23$ mag should be detected with a signal-to-noise ratio of at least $7$ using the observational setup adopted in this work. This consistency also provides an independent validation of the data reduction procedure described in Section \ref{sec:Observations and data reduction}.

\subsection{Minimum Mass Detection Limits}
\label{subsection: minimum mass}
The H-band apparent-magnitude detection limits were converted into minimum detectable companion masses using the ATMO evolutionary models available in \texttt{species}. For each WD, we computed the expected ERIS H-band apparent magnitudes of substellar companions over a grid of masses at the estimated total system age and the Gaia astrometric distance. The resulting mass--magnitude relation was then used to convert the H-band detection limits into minimum detectable companion masses at each projected separation. We used the three ATMO atmospheric grids: CEQ (chemical equilibrium), NEQ weak, and NEQ strong, which differ in their treatment of non-equilibrium chemistry induced by vertical mixing.

\begin{figure*}
    \centering
    \includegraphics[width=\textwidth]{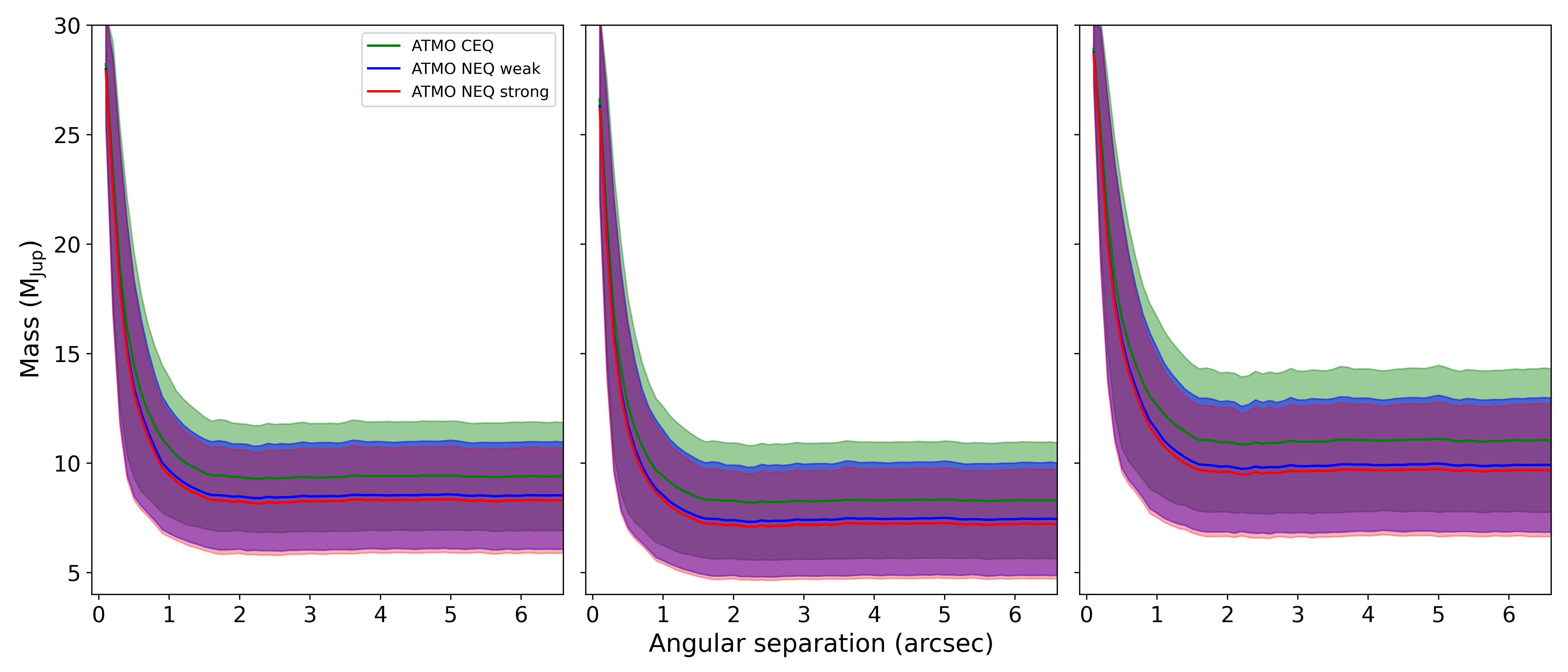}
    \caption{Minimum detectable companion masses derived from the contrast curves shown in Fig.~\ref{fig:contrast_curves} for the different ATMO models (see Section~\ref{subsection: minimum mass} for details). From left to right, the panels show the effect of WD age uncertainties on the minimum detectable companion masses, assuming the median age (left), the $-1\sigma$ age limit (center), and the $+1\sigma$ age limit (right). Overall, the different ATMO models and assumed ages yield a scatter of approximately $2\,M_\mathrm{Jup}$ in the minimum detectable masses.}
    \label{fig:_mass_detection_limit}
\end{figure*}

Figure~\ref{fig:_mass_detection_limit} shows the mean $5\sigma$ minimum mass detection limits obtained with the three ATMO model grids, adopting the minimum, median, and maximum total ages derived for the WD sample. As expected, the non-equilibrium chemistry (NEQ) models yield lower mass detection limits than the chemical equilibrium (CEQ) models, since vertical mixing enhances the emergent H-band flux of substellar objects at a given mass and age, making them brighter and therefore detectable at lower masses.

For the median total ages, the mean detection limit is approximately $8.4\,M_\mathrm{Jup}$ for both NEQ models, compared with $\sim9.4\,M_\mathrm{Jup}$ for the CEQ model. Adopting the minimum total ages, the mean detection limit decreases by approximately $1\,M_\mathrm{Jup}$, whereas adopting the maximum total ages increases it by about $1.5\,M_\mathrm{Jup}$. 

\subsection{Detection probability maps and survey completeness}

Using the minimum mass detection limits, we computed detection probability maps in the mass--semi-major-axis plane for each target with the Python package \texttt{EXO-DMC} \citep{Bonavita_2020}. For each grid point in the $(M,a)$ plane, the code generates 1000 orbital realizations. The orbital inclination is drawn from a distribution uniform in $\cos i$ (i.e., isotropically oriented orbits), while the eccentricity is assumed to follow a uniform distribution in the range $0 \leq e < 1$, given the current lack of observational constraints on the orbital properties of substellar companions around WDs. The remaining orbital elements are drawn from uniform distributions.

For each orbital realization, the projected separation is compared with the corresponding minimum mass detection limit, and the companion is considered detectable if its projected position falls within the $7''$ FoV and its mass exceeds the minimum detectable mass at that separation. The detection probability assigned to each $(M,a)$ grid point is then given by the fraction of detectable orbital realizations. The explored parameter space spans semi-major axes from 1 to 1000 au and companion masses from 1 to 75 $M_\mathrm{Jup}$.

\begin{figure}
    \centering

    \includegraphics[width=\columnwidth]
    {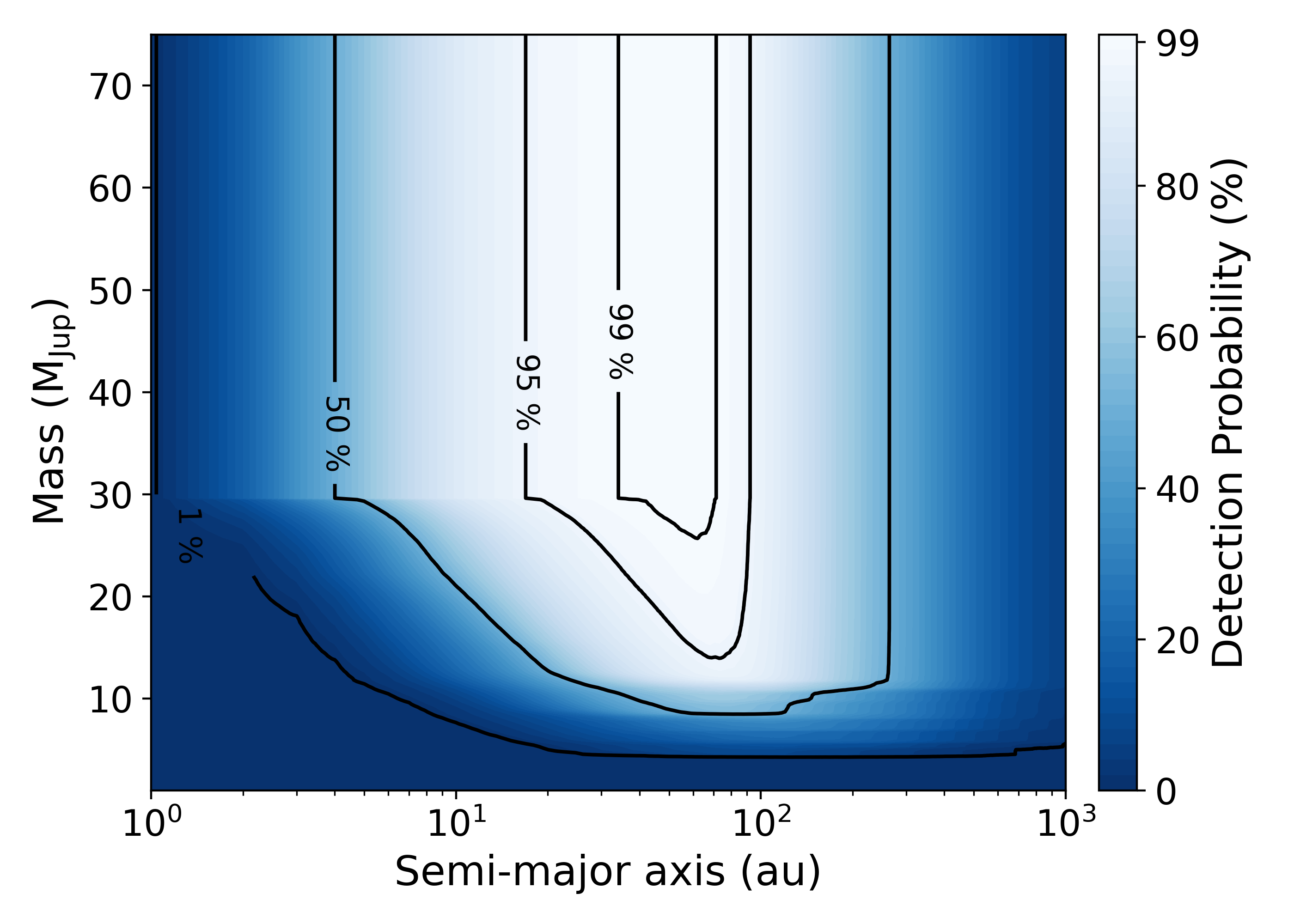}

    \par\vspace{0.05cm}

    \includegraphics[width=\columnwidth]
    {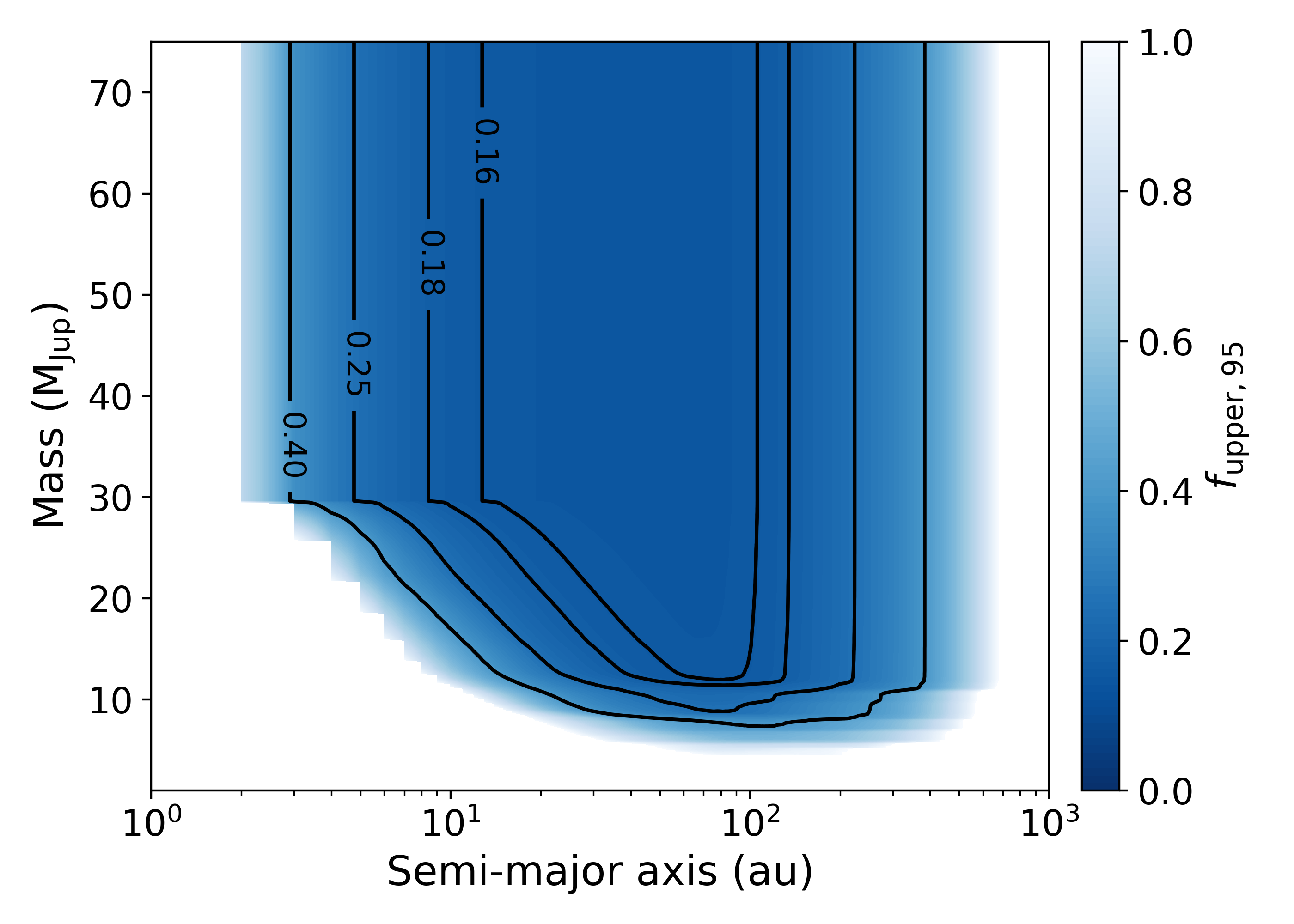}

    \caption{\textbf{Top}: Average detection probability as a function of companion mass and orbital semi-major axis. Our survey reaches a completeness above $50\%$ for companions more massive than about 12 Jupiter masses with semi-major axes between approximately 20 and 250 au. \textbf{Bottom}: $95\%$ confidence upper limit on the companion fraction across the mass–semi-major-axis plane assuming zero detections.}
    \label{fig:detection_map_and_f95}
\end{figure}

The top panel of figure \ref{fig:detection_map_and_f95} shows the sample-averaged detection probability map obtained with the ATMO NEQ strong model. We adopt this model as a representative case as non-equilibrium chemistry driven by vertical mixing is known to play an important role in the atmospheres of cool brown dwarfs, and models including this process have been shown to reproduce some observational properties better than chemical-equilibrium models \citep{Phillips_2020, Leggett2021}. Detection probabilities exceed 50 per cent for companions with masses above $12\,M_\mathrm{Jup}$ and semi-major axes between approximately $20$ and $250$\,au. At larger separations, the detection probability decreases as companions fall outside the $7''$ field of view. At separations below $\sim20$\,au, the halo of the WD PSF limits the sensitivity, restricting the detection of companions with masses below approximately $30\,M_\mathrm{Jup}$.

\subsection{Occurrence rate of substellar companions}

Using the detection probability map of each system, we estimated the $95\%$ confidence upper limit on the occurrence rate of substellar companions, $f_{\mathrm{upper},95}$, at each point in the mass--semi-major-axis plane, assuming that no confirmed companions were detected. Taking the detection of a companion as a binomial process, the probability of obtaining zero detections is
\begin{equation}
P(0)=0.05=\prod_{i=1}^{N=19}\left(1-f_{\mathrm{upper},95}\,C_i\right),
\label{eq:binomial}
\end{equation}
where $C_i$ is the detection probability map of the $i$th WD.

The lower panel of Fig. \ref{fig:detection_map_and_f95} shows the resulting 95\% confidence upper limits on the occurrence rate of substellar companions. The strongest constraint is obtained in the region where the survey is most sensitive, covering semi-major axes of approximately 50--100 au and companion masses above $\sim12\,M_\mathrm{Jup}$, where we derive $f_{\mathrm{upper},95}\approx0.16$. For more massive companions ($\gtrsim30\,M_\mathrm{Jup}$), a similar upper limit is achieved over a broader range of semi-major axes, extending from approximately $12$ to $100$\,au. The close correspondence between this region and the area of highest detection probability shown in the upper panel demonstrates that the occurrence-rate constraints are primarily driven by the survey completeness. Furthermore, even in the region of nearly complete sensitivity, the upper limit does not fall below $\sim16\%$, reflecting the finite sample size rather than the survey sensitivity.

\section{Discussion}

Having derived constraints on the occurrence rate of substellar companions around our sample of high-mass WDs, we now discuss the implications of these results. In particular, we examine how the potential presence of merger remnants in our sample may affect the inferred occurrence rate, and compare our results with previous surveys of substellar companions around WDs and their main-sequence progenitors. We also discuss our findings in the context of theoretical predictions for the formation and survival of substellar companions around intermediate-mass stars.

\subsection{Implications for the progenitor companion population}

Our primary motivation for surveying substellar companions to high-mass WDs is to constrain the population of such objects around their progenitor stars. Companions to main sequence stars 
with masses from $2-8$\,\Msun~are hard to detect, but they may provide key constraints on planet formation theories. Below four solar masses, core accretion is likely dominating planet formation, while for higher masses whereas gravitational instability and subsequent disc fragmentation are supposed to be the main formation mechanisms for substellar companions. 

Our survey of companions to WDs provided an upper limit on the occurrence rate of $16\%$ for substellar companions located at $50-100$~au. Assuming that all WDs in our sample formed through single star evolution, such wide companions were not affected by stellar tides and adiabatic expansion increased the orbital separation by a factor of $\sim3-6$ during the giant star phases of the progenitor. Therefore, the most sensitive region of our survey ($50-100$~au)~translates to separations of $10-30$~au~for companions around the WD progenitor 
stars.  Our observations therefore provide constraints on the population of massive substellar companions that formed at these intermediate separations and survived the subsequent evolution of their host stars. 

However, translating the observed upper limits into constraints on the original companion population is not entirely straightforward. Not all high-mass WDs necessarily descend from single stars, as some may have formed through binary mergers. In the following, we examine this limitation before we discuss possible constraints of our results on planet formation theories.  

\subsection{The impact of potential merger products on the companion occurrence rate}

Our survey constrains the occurrence of surviving substellar companions around high-mass WDs. However, one of its principal motivations is to investigate the companion population of their main-sequence progenitors, particularly intermediate-mass stars that evolved in isolation. This interpretation is complicated by the possibility that some of the observed WDs might have formed through binary mergers rather than single-star evolution.

Such merger remnants do not necessarily trace the same progenitor companion population as WDs formed through single-star evolution. In particular, the frequency and survival of substellar companions in close binary systems may differ from those around single stars \citep[e.g.,][]{Kraus2016, MoeKratter2021}. If merger remnants are present in the WD sample, the resulting occurrence-rate constraints may therefore not be representative of the population descended from single stars.

To quantify the potential impact of this contamination, we consider a limiting scenario in which merger remnants do not retain substellar companions. Adopting a mass-dependent merger formation probability, $f_{\rm merger}(M_\mathrm{WD})$, calculated for a small common envelope efficiency \citep{Zorotovic2010} from \citet{Temminketal_2020}, we statistically account for the unknown evolutionary origin of each target. 

For each of the $N=19$ WDs we drew $R=1000$ mass values uniformly between $\overline{M}_\mathrm{WD} - \delta M$ and $\overline{M}_\mathrm{WD}+\delta M$, where $\overline{M}_\mathrm{WD}$ is the median mass and $\delta M$ the mass dispersion reported in the last column of Table~\ref{tab:masses}. For WD $i$, the $r$-th mass realisation $M_{i,r}$ is assigned a merger flag $m_{i,r}=1$ with probability
$f_{\rm merger}(M_{i,r})$, and $m_{i,r}=0$ otherwise (i.e., corresponding to a WD formed through the single-star evolution). This yields an $N\times R$ array of merger flags, in which each realisation $r$ defines a subsample of putative non-mergers out of the 19 WDs.

For each realisation, we recomputed the $95\%$ confidence upper limit on the companion occurrence rate at every point in the mass--semi-major-axis plane, following Eq.~(\ref{eq:binomial}) but restricting the product to the putative non-merger subsample. We then calculated the average upper limit across the $R$ realisations. 

\begin{figure}
    \centering
    \includegraphics[width=\columnwidth]{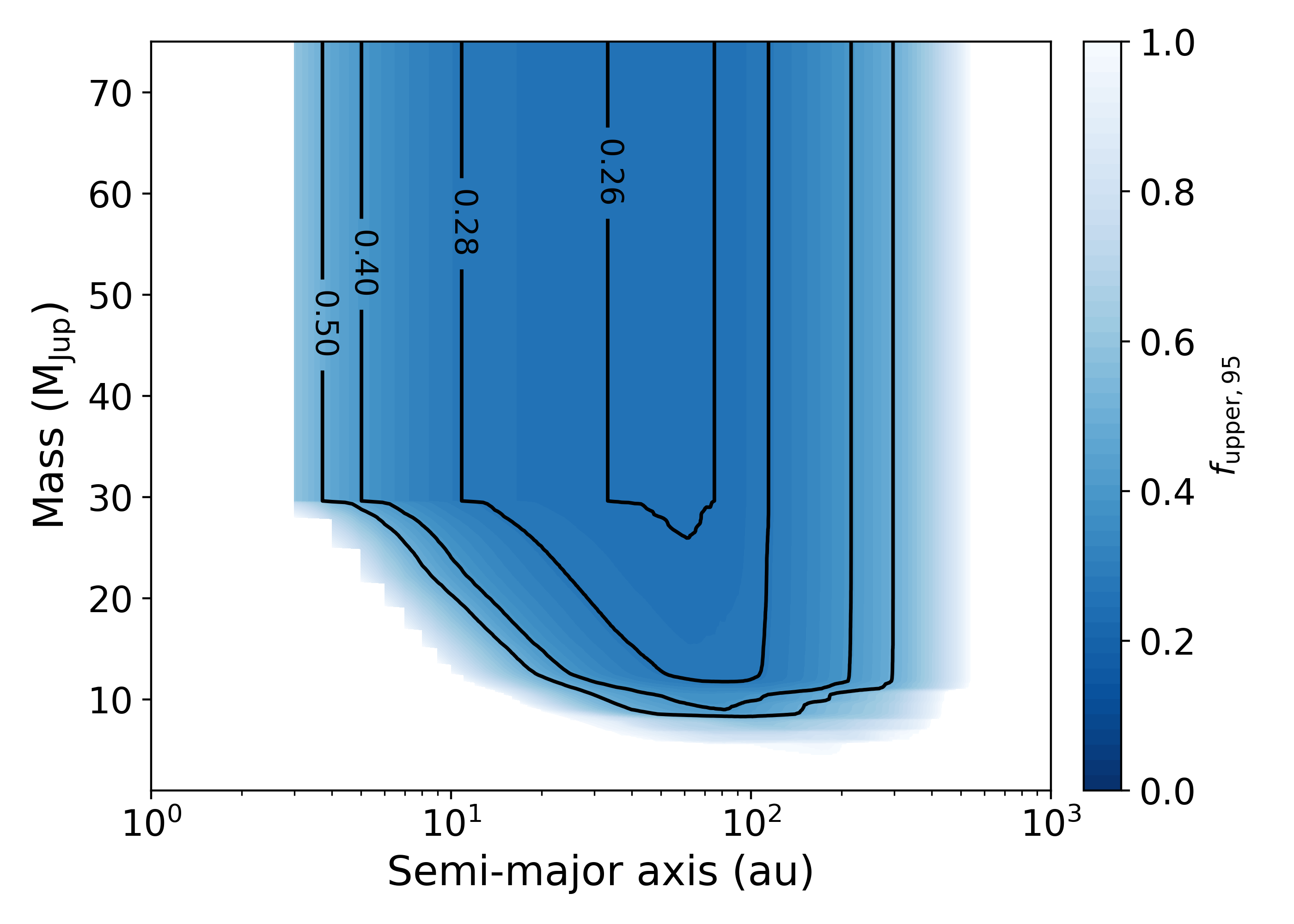}
    \caption{Same as bottom panel of Fig. \ref{fig:detection_map_and_f95}, but accounting for the potential presence of merger products in our WD sample. Overall, the upper limit on the companion occurrence rate, $f_{\rm upper,95}$, is less constraining by a factor of $\sim1.6$.}
    \label{fig:f95_merger_analysis}
\end{figure}

The number of potential mergers in our sample reduces the effective number of white dwarfs available to constrain the companion population descended from single stars. Under these assumptions, the $95\%$ upper limit increases from approximately $16\%$ to $26\%$ in the region of highest survey sensitivity (see Fig.\ref{fig:f95_merger_analysis}).

This result should not be interpreted as evidence that individual targets are merger remnants, nor as a correction to the measured companion occurrence rate. Instead, it illustrates the extent to which the unknown merger contribution may weaken constraints on the surviving companion population around WDs that descended from single-star progenitors.

\subsection{Comparison with previous companion surveys around WDs and their progenitors}

Previous direct-imaging surveys have established important constraints on the occurrence of planets and brown dwarfs around WDs and their main-sequence progenitors, although they differ in target selection, sensitivity, and orbital coverage. The DODO survey \citep{Hogan2009} used multi-epoch $J$-band imaging of 23 nearby WDs with masses between $0.49$--$1.07\,\mathrm{M}_\odot$ and found no confirmed comoving substellar companions. From its completeness analysis, they tentatively inferred that fewer than 5\% of WDs host companions with $T_{\rm eff}\gtrsim500$~K at projected separations of 60--200~au, at 90\% confidence. This result provides a useful benchmark for wide-orbit searches, although its temperature-based selection cannot be directly translated into a companion-mass threshold without adopting an evolutionary model and an age distribution.

Other imaging surveys provide complementary constraints. \citet{Debes2006} searched 13 nearby WDs with masses between $0.53$ and $0.97\,M_\odot$, ruling out companions more massive than approximately $14\,\mathrm{M_{Jup}}$ around eight targets at projected separations $\gtrsim19$~au and deriving an upper limit of 13\% on the frequency of companions more massive than $\sim10\,\mathrm{M_{Jup}}$. \citet{Farihi2005}, in a complementary near-infrared survey, searched 261 WDs at wide separations of approximately 100--5000~au and conducted a deeper search of 86 targets over $\sim50$--1100~au. Their observations were sensitive to companion masses down to approximately $0.05\,M_\odot$ and $0.02\,M_\odot$, respectively, and no brown-dwarf candidates were identified, leading to an estimated brown-dwarf companion fraction below 0.5\%. These results provide quantitative constraints on the general white-dwarf population, but the mass and separation ranges probed by each survey must be considered when comparing their occurrence limits with those derived from our survey.

The HST search of seven WDs in the Hyades cluster \citep{Brandner2021} provides a particularly relevant comparison in terms of progenitor mass. All seven targets descended from stars with initial masses above $2.8\,M_\odot$, and no planetary-mass companions were detected. Depending on the evolutionary models adopted, the observations reached detection limits of approximately $5$--$7\,\mathrm{M_{Jup}}$ or $9$--$12\,\mathrm{M_{Jup}}$ at projected separations corresponding to initial semimajor axes of $\gtrsim5$--$8$~au, before stellar mass loss. For reference, zero detections among seven targets correspond to a one-sided binomial upper limit of approximately $35\%$ at $95\%$ confidence. 
 
The census of confirmed substellar companions also highlights the limited number of systems involving massive WDs. A particularly relevant example is WD\,0837+185, a $\sim0.8\,M_\odot$ WD hosting a brown dwarf with a minimum mass of approximately $25\,\mathrm{M_{Jup}}$ in a close orbit of $\sim0.006$~au \citep{Casewell2012}. Its progenitor mass, estimated at $3.6\,M_\odot$, makes it relevant to the evolution of substellar companions around intermediate-mass stars. However, its compact orbit probes a different regime from the wide companions accessible to direct imaging. Conversely, several well-known systems with wide-orbit or planetary-mass companions, including GD\,165\,B \citep{Becklin1988} and WD\,0806$-$661\,B \citep{luhman_2012}, host WDs less massive than $0.7\,M_\odot$. Thus, while substellar companions are known across a range of white-dwarf systems, confirmed companions around WDs with $M_{\rm WD}\gtrsim0.7\,M_\odot$ remain particularly scarce at wide separations.

The results for main-sequence progenitors provide a further point of comparison. The Gemini NICI Planet-Finding Campaign \citep{nielsen2013} surveyed 70 young B- and A-type stars and used Monte Carlo simulations to constrain the frequency of wide-orbit giant planets. At 95\% confidence, fewer than 20\% of B- and A-type stars were found to host planets more massive than $4\,\mathrm{M_{Jup}}$ between 59 and 460~au, while fewer than 10\% could host planets more massive than $10\,\mathrm{M_{Jup}}$ between 38 and 650~au. The BEAST \citep{Delorme2024} survey provides a complementary result for young B stars: among the first 42 stars analysed, two substellar companions with masses near the deuterium-burning limit were identified, yielding a sensitivity-corrected occurrence rate of $11^{+7}_{-5}\%$. These surveys constrain the companion population before stellar mass loss, whereas our survey probes the remnants of intermediate-mass stars after their evolution. The corresponding orbital separations and companion luminosities are therefore not directly equivalent, but provide a useful reference for the occurrence of substellar companions around the progenitors of our WDs.

Overall, our results are broadly consistent with previous surveys of high-mass WDs and their main-sequence progenitors, which have generally found a low occurrence of substellar companions. Among WD surveys, our upper limit of $f_{\rm upper,95}\approx16\%$ for companions more massive than $\sim12\,\mathrm{M_{Jup}}$ at separations of 50–100 au is compatible with the constraints from DODO, as well as with the non-detections reported by \citet{Farihi2005}, \citet{Debes2006} and \citet{Brandner2021}. Similarly, the NICI survey and the BEAST survey constrain the frequency of substellar companions around B- and A-type stars before stellar mass loss, with results that are compatible with the upper limits derived here. Accounting for potential merger products weakens our upper limit to $\sim26\%$, but does not alter this overall picture. Nevertheless, these comparisons should be interpreted with caution, since the surveys probe different companion-mass and orbital-separation ranges, as well as different stages of stellar evolution. Thus, while our non-detection does not reveal any clear discrepancy with previous observational constraints, it does not demonstrate that the occurrence rates are equivalent across these populations.

\subsection{Comparison with theoretical predictions}

Population-synthesis models provide a framework for connecting the formation mechanisms of substellar companions to the population that may survive around WDs. This is particularly relevant for intermediate-mass stars, for which core accretion may become less efficient as disc lifetimes shorten and photoevaporation becomes more important, potentially increasing the relative importance of GI \citep{KennedyKenyon2008,KratterLodato2016, Ronco2024}. The predicted companion populations depend on the formation mechanism and on the subsequent accretion, orbital migration, and dynamical evolution of the objects.

The DIPSY population synthesis of companions formed through disc instability predicts a dependence of companion properties on the mass of the host star \citep{Schib2025}. As shown in fig.~9 of \citet{Schib2025}, the median semi-major axis of companions increases with stellar mass, following $a/\mathrm{au}\simeq550(M_\star/M_\odot)^{0.6}$. For host-star masses of $2$--$5\,M_\odot$, this relation corresponds to median semi-major axes of approximately $800$--$1400$~au. DIPSY also predicts a population dominated by brown dwarfs, with planetary-mass companions being particularly scarce at separations $\lesssim~100$\,au. These results remain broadly similar across most of their simulation variants, although the predicted number of surviving companions can vary with the adopted initial disc properties and the treatment of fragment interactions. For companions that survive post-main-sequence evolution, assuming isotropic and adiabatic stellar mass loss, the semi-major axis increases approximately by a factor of $M_{\rm MS}/M_{\rm WD}$, where $M_{\rm MS}$ and $M_{\rm WD}$ are the progenitor and white-dwarf masses, respectively. Companions initially located at separations exceeding several hundred au would therefore move to even wider orbits with separations larger than $1000$~au. 

Core accretion provides a complementary formation channel, with population-synthesis models predicting companion populations shaped by solid-core growth, gas accretion, and orbital migration \citep{Mordasini2009}. Using pebble-driven core-accretion models for host stars of $1$--$2.4\,M_\odot$, \citet{Johnston2024} and, in more detail, Ronco et al. (2026, A\&A submitted) showed that giant-planet formation depends strongly on the initial disc properties, embryo location, time of embryo formation, migration, and the treatment of photoevaporation. In these models, planets can form at different locations and may even remain at $\gappr1$--$30$\,au because of thermal torques. 
However, giant-planet formation becomes challenging around stars more massive than $\sim3.5$~\Msun. 

Companions formed on such relatively compact orbits are more susceptible to tidal interactions and engulfment during post-main-sequence evolution than objects formed at wider separations.
The importance of this subsequent evolution is illustrated by \citet{Mauch-Soriano_2026}, who predict that fewer than $\sim3\%$ of WDs host surviving substellar, mostly planetary mass, 
companions. The surviving companions have semi-major axes of $\approx3$--$24$\,au, and their simulations explicitly account for stellar tides and mass loss. Engulfed companions that may survive common envelope evolution are extremely rare and negligible for population studies \citep{mirandaetal26-1}.  

In summary, formation through core accretion predicts a population 
of close in planetary companions with separations ranging from a few to a few dozens of au 
around WDs descending from stars less massive than $\sim3.5$~\Msun. In addition, for WD progenitor masses ranging from $1-8$~\Msun, and nearly independent of stellar mass, mostly brown dwarf companions are predicted at separations exceeding $1000$~au. 

These very recent predictions, that were not available when we started our observing campaign,  have interesting implications for our survey. As shown in Fig. \ref{fig:detection_map_and_f95}, our survey is not sensitive to planetary mass companions at short separations so we are unable to detect most surviving giant planets predicted to form through core accretion. At the same time, the field of view of ERIS is too small to detect companions at separations exceeding several hundred au. Therefore, the population of brown dwarfs at separations exceeding thousands of au, predicted to form around WD progenitors through disc instability, will escape detection by ERIS as well. It might thus well be that we observed a parameter space that is not sensitive to existing substellar companions around WDs. 
While we note that theoretical predictions and population synthesis of planet formation are notoriously uncertain and therefore observational constraints of virtually the entire parameter space are useful, a legitimate question is which type of observing strategy is most promising for providing constraints on planet formation from observing companions to WDs.  

As shown in Mauch-Soriano et al. (2026, A\&A submitted), through the low occurrence rates of close  companions even ongoing JWST programs are unlikely to find large numbers of close companions around white dwarfs descending from $4-8$~\Msun~stars. The large field of view of JWST, however, should be able to detect substellar companions that formed through disc instabilities. As the main sequence lifetime of their progenitors is short, and their companions therefore relatively hot, massive WDs as targeted here represent an ideal population to confront the prediction of substellar companion formation through disc instabilities. 

\section{Conclusions}

We have presented a high-contrast imaging survey of 19 high-mass WDs using ERIS/NIX at the VLT, aimed at constraining the occurrence rate of substellar companions around the remnants of intermediate-mass stars. We found no confirmed companions in our sample. Based on our detection sensitivities, we derived a $95\%$ confidence upper limit of approximately $16\%$ on the occurrence rate of companions more massive than $\sim12\,\mathrm{M_{Jup}}$ at semi-major axes of $\approx50$--$100$\,au~which translates to separations of $\sim10-30$~au around their progenitor stars. 

We also investigated how the potential presence of merger products in our sample could affect these constraints, which is important as the existence of mergers in our sample of WDs affects our constraints on the population of companions around WD progenitors. By statistically accounting for potential merger products using population-synthesis predictions, we find that the $95\%$ upper limit on the companion occurrence rate increases from approximately $16\%$ to $26\%$. This result illustrates how uncertainties in the merger fraction can affect constraints on the occurrence of substellar companions around WD progenitors, under the assumption that merger products do not retain surviving substellar companions.

While ERIS limited field of view prevents the detection of wide companions predicted by disc instability models, our survey still provides new observational constraints on the population of substellar companions around high-mass WDs. Given the larger field of view and the higher sensitivity, a dedicated JWST/MIRI survey of high-mass WDs could provide key constraints on planet formation through disc instabilities.

\begin{acknowledgements}
FLV acknowledges support from ANID
FONDECYT Postdoctoral Grant No. 3250464. MRS thanks for support from the research unit FOR2990 (eRO-STEP: Stellar endpoints with eROSITA) funded by the Deutsche Forschungsgemeinschaft and the internal research project PI\_LIR\_26\_07 at USM. AMS acknowledges support from Agencia Nacional de Investigación y Desarrollo (ANID) Becas/Doctorado Nacional 21263654. 

This publication makes use of data products from the Two Micron All Sky Survey, which is a joint project of the University of Massachusetts and the Infrared Processing and Analysis Center/California Institute of Technology, funded by the National Aeronautics and Space Administration and the National Science Foundation. This publication makes use of data products from the Spectro-Photometer for the History of the Universe, Epoch of Reionization and Ices Explorer (SPHEREx), which is a joint project of the Jet Propulsion Laboratory and the California Institute of Technology, and is funded by the National Aeronautics and Space Administration. Based on observations made with ESO Telescopes at the Paranal Observatory under programmes 111.251A.001, 113.26EH.001, 113.26EH.002, 114.27BT.001 and 117.2AYA.001. 

\end{acknowledgements}

%
\bibliographystyle{aa} 
\bibliography{WD_eris} 







   
  



\begin{appendix}




\onecolumn
\section{Tables}

\begin{table}
\centering
\caption{
Astrometric, photometric, and atmospheric parameters of the WDs in our sample. The coordinates (J2000), spectral types, and distances were collected from the Montreal WD Database \citep{Dufour2017} and SIMBAD \citep{simbad_ref}. The $H$-band magnitudes are taken from 2MASS, except for the three targets with poor-quality 2MASS photometry, for which synthetic magnitudes were derived using SPHEREx spectrophotometry (shown in bold). Effective temperatures ($T_{\rm eff}$), surface gravities ($\log g$), and WD masses ($M_{\rm WD}$) are adopted from \cite{Vincent_2024}.}
\label{tab:WD_sample}

\begin{tabular}{lcccccccc}

\hline

Simbad ID &
\begin{tabular}[c]{@{}c@{}}RA\\ (J2000)\end{tabular} &
\begin{tabular}[c]{@{}c@{}}Dec.\\ (J2000)\end{tabular} &
SpT &
\begin{tabular}[c]{@{}c@{}}Distance\\ (pc)\end{tabular} &
\begin{tabular}[c]{@{}c@{}}$H$\\ (mag)\end{tabular} &
\begin{tabular}[c]{@{}c@{}}$T_\mathrm{eff}$\\ (K)\end{tabular} &
\begin{tabular}[c]{@{}c@{}}$\log g$\\ (dex)\end{tabular} &
\begin{tabular}[c]{@{}c@{}}Mass\\ ($\mathrm{M}_\odot$)\end{tabular}

\\
\noalign{\smallskip}
\hline

\noalign{\smallskip}

EGGR 140 &
20:44:21.98 &
-68:05:25.01 &
DA &
\errnum{19.57}{0.01} &
\errnum{13.81}{0.04} &
\errnum{16840}{97} &
\errnum{8.487}{0.004} &
\errnum{0.920}{0.003}
\\
\noalign{\smallskip}

EGGR 155 &
22:49:06.15 &
+22:36:32.94 &
DA &
\errnum{17.83}{0.01} &
\errnum{14.32}{0.05} &
\errnum{10342}{58} &
\errnum{8.641}{0.007} &
\errnum{1.008}{0.005}
\\
\noalign{\smallskip}

UCAC4 230-183848 &
19:49:25.99 &
-44:05:12.83 &
DA &
\errnum{57.12}{0.14} &
\errnum{15.49}{0.10} &
\errnum{27219}{363} &
\errnum{8.499}{0.013} &
\errnum{0.94}{0.01}
\\
\noalign{\smallskip}

LP 873-78 &
21:21:21.73 &
-25:57:18.10 &
DA &
\errnum{24.52}{0.03} &
\errnum{13.46}{0.03} &
\errnum{19920}{181} &
\errnum{8.115}{0.007} &
\errnum{0.691}{0.006}
\\
\noalign{\smallskip}

PG 1350-090 &
13:53:15.92 &
-09:16:39.11 &
DA &
\errnum{19.71}{0.01} &
\errnum{14.08}{0.01} &
\errnum{8727}{41} &
\errnum{8.295}{0.007} &
\errnum{0.783}{0.006}
\\
\noalign{\smallskip}

PHL 1716 &
21:43:24.08 &
-06:59:49.46 &
DA &
\errnum{18.15}{0.01} &
\errnum{14.18}{0.04} &
\errnum{8970}{30} &
\errnum{8.438}{0.005} &
\errnum{0.878}{0.005}
\\
\noalign{\smallskip}



V* CD Oct &
22:04:18.65 &
-75:13:24.88 &
DA &
\errnum{19.03}{0.01} &
\errnum{14.67}{0.07} &
\errnum{8908}{34} &
\errnum{8.633}{0.005} &
\errnum{1.002}{0.004}
\\
\noalign{\smallskip}

WD 2051-208 &
20:54:42.88 &
-20:39:25.76 &
DA &
\errnum{31.28}{0.03} &
\errnum{15.67}{0.11} &
\errnum{20236}{164} &
\errnum{9.049}{0.006} &
\errnum{1.222}{0.003}
\\
\noalign{\smallskip}

WD 2151-307 &
21:54:53.43 &
-30:29:18.88 &
DA &
\errnum{69.07}{0.21} &
\errnum{15.44}{0.10} &
\errnum{28999}{307} &
\errnum{8.29}{0.01} &
\errnum{0.815}{0.009}
\\
\noalign{\smallskip}

GD 1706 &
23:53:03.77 &
-24:32:03.50 &
DA &
\errnum{87.75}{0.37} &
\errnum{\textbf{16.06}}{\textbf{0.03}} &
\errnum{30193}{575} &
\errnum{8.432}{0.018} &
\errnum{0.902}{0.015}
\\
\noalign{\smallskip}

LAWD 83 &
21:13:18.84 &
-81:49:18.96 &
DA &
\errnum{16.18}{0.004} &
\errnum{13.45}{0.03} &
\errnum{9513}{32} &
\errnum{8.170}{0.005} &
\errnum{0.703}{0.005}
\\
\noalign{\smallskip}

L 426-26 &
21:38:49.29 &
-40:41:26.93 &
DA &
\errnum{32.43}{0.03} &
\errnum{14.04}{0.04} &
\errnum{19827}{163} &
\errnum{8.103}{0.007} &
\errnum{0.683}{0.005}
\\
\noalign{\smallskip}

RX J1847.9-2219 &
18:47:56.62 &
-22:19:38.36 &
DA &
\errnum{51.41}{0.11} &
\errnum{\textbf{14.53}}{\textbf{0.01}} &
\errnum{39713}{1475} &
\errnum{8.423}{0.020} &
\errnum{0.909}{0.016}
\\
\noalign{\smallskip}

WD 0024-556 &
00:26:40.17 &
-55:24:51.37 &
DA &
\errnum{24.05}{0.01} &
\errnum{15.23}{0.10} &
\errnum{10058}{44} &
\errnum{8.665}{0.006} &
\errnum{1.022}{0.005}
\\
\noalign{\smallskip}

WD 0033+013 &
00:35:35.68 &
01:53:02.82 &
DA &
\errnum{31.84}{0.04} &
\errnum{15.52}{0.09} &
\errnum{10505}{74} &
\errnum{8.662}{0.009} &
\errnum{1.021}{0.007}
\\
\noalign{\smallskip}

WD 0136-201 &
01:38:32.26 &
-19:54:43.99 &
DA &
\errnum{24.38}{0.02} &
\errnum{15.10}{0.09} &
\errnum{8319}{39} &
\errnum{8.507}{0.007} &
\errnum{0.922}{0.006}
\\
\noalign{\smallskip}

EC 04426-3523 &
04:44:27.32 &
-35:17:46.25 &
DB &
\errnum{59.09}{0.11} &
\errnum{\textbf{16.18}}{\textbf{0.03}} &
\errnum{13894}{130} &
\errnum{8.246}{0.010} &
\errnum{0.740}{0.009}
\\
\noalign{\smallskip}

GALEX J081835.6-183243 &
08:18:35.69 &
-18:32:43.29 &
DB &
\errnum{87.66}{0.35} &
\errnum{\textbf{16.84}}{\textbf{0.05}}&
\errnum{15524}{271} &
\errnum{8.226}{0.019} &
\errnum{0.729}{0.017}
\\
\noalign{\smallskip}

UCAC4 460-008869 &
05:09:41.03 &
01:53:08.40 &
DB &
\errnum{64.10}{0.19} &
\errnum{15.49}{0.11} &
\errnum{21586}{270} &
\errnum{8.327}{0.008} &
\errnum{0.801}{0.007}
\\
\noalign{\smallskip}

\hline
\\
\end{tabular}
\end{table}

\begin{table}
\renewcommand{\arraystretch}{2}
\caption{Total ages ($\tau_{\rm Tot}$) of the WDs in our sample, in Gyr, derived using the six combinations of initial-to-final mass relations (IFMRs) and WD cooling tracks considered in this work. The columns list the age estimates obtained with the Montreal \citep[B20, ][]{Bedard2020} and La Plata \citep[LPCODE, ][]{Camisassa_2016,Camisassa_2017, Camisassa_2019} cooling tracks, combined with the IFMRs from \citet{Cummings_2018} (CP18), \citet{Cunningham_2024} (C24), and \citet{Miller_2026} (M26). The final adopted age for each target is given in the last column and corresponds to the median of the six estimates, with uncertainties spanning their minimum and maximum values.}
\label{tab:total_ages}
\centering

\begin{tabular}{lccc c ccc c}
\hline

&
\multicolumn{3}{c}{B20 cooling tracks} &
&
\multicolumn{3}{c}{LPCODE cooling tracks}
\\

\cline{2-4}
\cline{6-8}

Name &
$\mathrm{\tau^{CP18}_{Tot}}$ &
$\mathrm{\tau^{C24}_{Tot}}$  &
$\mathrm{\tau^{M26}_{Tot}}$  &
&
$\mathrm{\tau^{CP18}_{Tot}}$ &
$\mathrm{\tau^{CP24}_{Tot}}$ &
$\mathrm{\tau^{M26}_{Tot}}$&

$\overline{\tau}_\mathrm{Tot}$
\\

\hline

EGGR 140 & $0.51^{+0.04}_{-0.03}$ & $0.45^{+0.03}_{-0.02}$ & $0.56^{+0.04}_{-0.04}$ &  & $0.54^{+0.05}_{-0.04}$ & $0.47^{+0.03}_{-0.02}$ & $0.59^{+0.04}_{-0.04}$ &$0.53^{+0.10}_{-0.10}$  \\
EGGR 155 & $1.98^{+0.11}_{-0.12}$ & $1.96^{+0.12}_{-0.12}$ & $2.01^{+0.11}_{-0.11}$ &  & $1.80^{+0.08}_{-0.07}$ & $1.78^{+0.08}_{-0.08}$ & $1.84^{+0.08}_{-0.07}$ &$1.90^{+0.22}_{-0.25}$  \\
UCAC4 230-183848 & $0.23^{+0.04}_{-0.03}$ & $0.18^{+0.02}_{-0.01}$ & $0.27^{+0.04}_{-0.04}$ &  & $0.25^{+0.05}_{-0.04}$ & $0.19^{+0.03}_{-0.01}$ & $0.29^{+0.04}_{-0.04}$ &$0.24^{+0.10}_{-0.07}$  \\
LP 873-78 & $1.10^{+0.50}_{-0.35}$ & $0.99^{+0.47}_{-0.30}$ & $1.45^{+0.57}_{-0.53}$ &  & $1.23^{+0.58}_{-0.41}$ & $1.10^{+0.48}_{-0.34}$ & $1.49^{+0.73}_{-0.51}$ &$1.17^{+0.89}_{-0.47}$  \\
PG 1350-090 & $1.83^{+0.10}_{-0.08}$ & $1.75^{+0.11}_{-0.09}$ & $1.89^{+0.14}_{-0.09}$ &  & $1.98^{+0.12}_{-0.09}$ & $1.91^{+0.12}_{-0.10}$ & $2.06^{+0.24}_{-0.11}$ &$1.90^{+0.29}_{-0.24}$  \\
PHL 1716 & $2.08^{+0.14}_{-0.14}$ & $1.97^{+0.14}_{-0.14}$ & $2.13^{+0.14}_{-0.14}$ &  & $1.99^{+0.09}_{-0.08}$ & $1.89^{+0.09}_{-0.08}$ & $2.04^{+0.09}_{-0.08}$ &$2.01^{+0.26}_{-0.26}$  \\
V* CD Oct & $2.84^{+0.11}_{-0.12}$ & $2.82^{+0.12}_{-0.12}$ & $2.87^{+0.11}_{-0.11}$ &  & $2.59^{+0.08}_{-0.08}$ & $2.56^{+0.09}_{-0.09}$ & $2.63^{+0.08}_{-0.08}$ &$2.73^{+0.26}_{-0.28}$  \\
WD 2051-208 & $0.64^{+0.05}_{-0.05}$ & $0.66^{+0.05}_{-0.05}$ & $0.65^{+0.05}_{-0.05}$ &  & $0.65^{+0.03}_{-0.03}$ & $0.67^{+0.03}_{-0.03}$ & $0.66^{+0.03}_{-0.03}$ &$0.66^{+0.06}_{-0.06}$  \\
WD 2151-307 & $0.36^{+0.06}_{-0.05}$ & $0.27^{+0.08}_{-0.06}$ & $0.41^{+0.08}_{-0.06}$ &  & $0.39^{+0.07}_{-0.06}$ & $0.30^{+0.10}_{-0.07}$ & $0.44^{+0.09}_{-0.07}$ &$0.37^{+0.15}_{-0.17}$  \\
GD 1706 & $0.23^{+0.05}_{-0.04}$ & $0.16^{+0.03}_{-0.02}$ & $0.28^{+0.04}_{-0.05}$ &  & $0.26^{+0.05}_{-0.05}$ & $0.18^{+0.04}_{-0.03}$ & $0.31^{+0.05}_{-0.04}$ &$0.25^{+0.11}_{-0.10}$  \\
LAWD 83 & $1.78^{+0.48}_{-0.26}$ & $1.68^{+0.44}_{-0.23}$ & $2.19^{+0.43}_{-0.53}$ &  & $1.92^{+0.46}_{-0.31}$ & $1.82^{+0.47}_{-0.26}$ & $2.30^{+0.51}_{-0.53}$ &$1.87^{+0.86}_{-0.41}$  \\
L 426-26 & $1.21^{+0.58}_{-0.40}$ & $1.08^{+0.48}_{-0.34}$ & $1.48^{+0.71}_{-0.51}$ &  & $1.35^{+0.71}_{-0.48}$ & $1.21^{+0.54}_{-0.40}$ & $1.57^{+0.87}_{-0.54}$ &$1.28^{+0.99}_{-0.54}$  \\
RX J1847.9-2219 & $0.20^{+0.05}_{-0.04}$ & $0.13^{+0.03}_{-0.02}$ & $0.24^{+0.05}_{-0.05}$ &  & $0.22^{+0.05}_{-0.05}$ & $0.15^{+0.04}_{-0.03}$ & $0.27^{+0.05}_{-0.05}$ &$0.21^{+0.11}_{-0.10}$  \\
WD 0024-556 & $2.21^{+0.10}_{-0.12}$ & $2.20^{+0.11}_{-0.12}$ & $2.24^{+0.10}_{-0.11}$ &  & $1.98^{+0.10}_{-0.08}$ & $1.96^{+0.11}_{-0.08}$ & $2.02^{+0.10}_{-0.08}$ &$2.11^{+0.23}_{-0.26}$  \\
WD 0033+013 & $1.97^{+0.10}_{-0.12}$ & $1.96^{+0.11}_{-0.12}$ & $1.99^{+0.10}_{-0.11}$ &  & $1.77^{+0.10}_{-0.08}$ & $1.75^{+0.11}_{-0.08}$ & $1.81^{+0.10}_{-0.08}$ &$1.88^{+0.21}_{-0.25}$  \\
WD 0136-201 & $2.85^{+0.13}_{-0.14}$ & $2.78^{+0.14}_{-0.15}$ & $2.91^{+0.13}_{-0.14}$ &  & $2.58^{+0.15}_{-0.13}$ & $2.49^{+0.16}_{-0.13}$ & $2.64^{+0.15}_{-0.14}$ &$2.71^{+0.33}_{-0.37}$  \\
EC 04426-3523 & $0.97^{+0.28}_{-0.13}$ & $0.89^{+0.23}_{-0.13}$ & $1.18^{+0.60}_{-0.26}$ &  & $1.04^{+0.32}_{-0.16}$ & $0.96^{+0.27}_{-0.14}$ & $1.35^{+0.48}_{-0.36}$ & $1.00^{+0.94}_{-0.24}$  \\
GALEX J081835.6-183243 & $0.92^{+0.33}_{-0.17}$ & $0.84^{+0.28}_{-0.15}$ & $1.27^{+0.45}_{-0.40}$ &  & $0.99^{+0.39}_{-0.19}$ & $0.91^{+0.33}_{-0.18}$ & $1.40^{+0.40}_{-0.48}$ & $0.96^{+0.90}_{-0.27}$  \\
UCAC4 460-008869 & $0.48^{+0.07}_{-0.06}$ & $0.39^{+0.10}_{-0.07}$ & $0.53^{+0.09}_{-0.07}$ &  & $0.50^{+0.08}_{-0.06}$ & $0.41^{+0.10}_{-0.08}$ & $0.55^{+0.11}_{-0.07}$ & $0.49^{+0.16}_{-0.17}$  \\

\\
\hline
\end{tabular}

\end{table}

\begin{table}
\renewcommand{\arraystretch}{2}
\caption{Progenitor ($M_{\rm prog}$) and white dwarf ($M_{\rm WD}$) masses, in $\mathrm{M}_\odot$, derived using the three IFMRs and two WD cooling tracks described in Table \ref{tab:total_ages}. The first three columns list the progenitor masses obtained with each IFMR, followed by the adopted progenitor mass. The next two columns show the WD masses derived using each cooling track, while the final column lists the adopted WD mass.}
\label{tab:masses}
\centering

\begin{tabular}{lcccc c ccc}
\hline



Name &
$M^\mathrm{CP18}_\mathrm{prog}$ &
$M^\mathrm{C24}_\mathrm{prog}$  &
$M^\mathrm{M26}_\mathrm{prog}$  &
$\overline{M}_\mathrm{prog}$ &
&
$M^\mathrm{B20}_\mathrm{WD}$ &
$M^\mathrm{LP}_\mathrm{WD}$ &
$\overline{M}_\mathrm{WD}$
\\

\hline

EGGR 140 & $4.19^{+0.42}_{-0.37}$ & $4.94^{+0.31}_{-0.42}$ & $3.86^{+0.34}_{-0.25}$ & $4.19^{+1.05}_{-0.58}$ &  & $0.92^{+0.02}_{-0.02}$ & $0.90^{+0.02}_{-0.02}$ & $0.91^{+0.03}_{-0.03}$ \\
EGGR 155 & $5.22^{+0.42}_{-0.42}$ & $5.63^{+0.21}_{-0.21}$ & $4.73^{+0.44}_{-0.44}$ & $5.22^{+0.63}_{-0.93}$ &  & $1.01^{+0.02}_{-0.02}$ & $0.99^{+0.02}_{-0.02}$ & $1.00^{+0.03}_{-0.03}$ \\
UCAC4 230-183848 & $4.41^{+0.42}_{-0.41}$ & $5.14^{+0.25}_{-0.38}$ & $4.02^{+0.37}_{-0.28}$ & $4.41^{+0.98}_{-0.67}$ &  & $0.94^{+0.02}_{-0.02}$ & $0.92^{+0.02}_{-0.02}$ & $0.93^{+0.03}_{-0.03}$ \\
LP 873-78 & $2.29^{+0.37}_{-0.41}$ & $2.38^{+0.36}_{-0.39}$ & $2.03^{+0.43}_{-0.31}$ & $2.29^{+0.45}_{-0.57}$ &  & $0.69^{+0.02}_{-0.02}$ & $0.68^{+0.02}_{-0.02}$ & $0.68^{+0.02}_{-0.02}$ \\
PG 1350-090 & $3.13^{+0.21}_{-0.24}$ & $3.37^{+0.42}_{-0.35}$ & $2.98^{+0.21}_{-0.27}$ & $3.13^{+0.66}_{-0.42}$ &  & $0.78^{+0.02}_{-0.02}$ & $0.77^{+0.02}_{-0.02}$ & $0.78^{+0.03}_{-0.03}$ \\
PHL 1716 & $3.80^{+0.36}_{-0.26}$ & $4.46^{+0.42}_{-0.46}$ & $3.54^{+0.23}_{-0.20}$ & $3.80^{+1.09}_{-0.46}$ &  & $0.87^{+0.02}_{-0.02}$ & $0.86^{+0.02}_{-0.02}$ & $0.87^{+0.03}_{-0.03}$ \\
V* CD Oct & $5.13^{+0.42}_{-0.42}$ & $5.59^{+0.21}_{-0.21}$ & $4.64^{+0.43}_{-0.42}$ & $5.13^{+0.67}_{-0.92}$ &  & $1.00^{+0.02}_{-0.02}$ & $0.98^{+0.02}_{-0.02}$ & $0.99^{+0.03}_{-0.03}$ \\
WD 2051-208 & $7.73^{+0.28}_{-0.35}$ & $6.94^{+0.19}_{-0.19}$ & $7.45^{+0.39}_{-0.41}$ & $7.45^{+0.56}_{-0.70}$ &  & $1.22^{+0.01}_{-0.01}$ & $1.19^{+0.01}_{-0.01}$ & $1.20^{+0.03}_{-0.03}$ \\
WD 2151-307 & $3.35^{+0.22}_{-0.20}$ & $3.75^{+0.44}_{-0.38}$ & $3.19^{+0.20}_{-0.19}$ & $3.35^{+0.84}_{-0.35}$ &  & $0.82^{+0.02}_{-0.02}$ & $0.80^{+0.02}_{-0.02}$ & $0.81^{+0.03}_{-0.03}$ \\
GD 1706 & $4.02^{+0.39}_{-0.33}$ & $4.75^{+0.35}_{-0.43}$ & $3.72^{+0.30}_{-0.22}$ & $4.02^{+1.08}_{-0.52}$ &  & $0.90^{+0.02}_{-0.02}$ & $0.88^{+0.02}_{-0.02}$ & $0.89^{+0.03}_{-0.03}$ \\
LAWD 83 & $2.41^{+0.35}_{-0.39}$ & $2.51^{+0.36}_{-0.39}$ & $2.10^{+0.45}_{-0.31}$ & $2.41^{+0.46}_{-0.62}$ &  & $0.70^{+0.02}_{-0.02}$ & $0.69^{+0.02}_{-0.02}$ & $0.69^{+0.02}_{-0.02}$ \\
L 426-26 & $2.20^{+0.38}_{-0.41}$ & $2.30^{+0.36}_{-0.39}$ & $1.97^{+0.43}_{-0.31}$ & $2.20^{+0.46}_{-0.54}$ &  & $0.68^{+0.02}_{-0.02}$ & $0.67^{+0.02}_{-0.02}$ & $0.68^{+0.02}_{-0.02}$ \\
RX J1847.9-2219 & $4.11^{+0.40}_{-0.35}$ & $4.84^{+0.33}_{-0.42}$ & $3.78^{+0.32}_{-0.23}$ & $4.11^{+1.07}_{-0.55}$ &  & $0.91^{+0.02}_{-0.02}$ & $0.89^{+0.02}_{-0.02}$ & $0.90^{+0.03}_{-0.03}$ \\
WD 0024-556 & $5.39^{+0.42}_{-0.42}$ & $5.72^{+0.21}_{-0.21}$ & $4.90^{+0.44}_{-0.44}$ & $5.39^{+0.54}_{-0.93}$ &  & $1.02^{+0.02}_{-0.02}$ & $1.00^{+0.02}_{-0.02}$ & $1.01^{+0.03}_{-0.03}$ \\
WD 0033+013 & $5.37^{+0.42}_{-0.41}$ & $5.72^{+0.21}_{-0.21}$ & $4.88^{+0.45}_{-0.44}$ & $5.37^{+0.55}_{-0.93}$ &  & $1.02^{+0.02}_{-0.02}$ & $1.00^{+0.02}_{-0.02}$ & $1.01^{+0.03}_{-0.03}$ \\
WD 0136-201 & $4.33^{+0.42}_{-0.39}$ & $5.07^{+0.28}_{-0.40}$ & $3.96^{+0.36}_{-0.26}$ & $4.33^{+1.01}_{-0.63}$ &  & $0.93^{+0.02}_{-0.02}$ & $0.91^{+0.02}_{-0.02}$ & $0.92^{+0.03}_{-0.03}$ \\
EC 04426-3523 & $2.77^{+0.27}_{-0.36}$ & $2.91^{+0.34}_{-0.36}$ & $2.48^{+0.37}_{-0.52}$ & $2.77^{+0.48}_{-0.81}$ &  & $0.74^{+0.02}_{-0.02}$ & $0.73^{+0.02}_{-0.02}$ & $0.73^{+0.02}_{-0.02}$ \\
GALEX J081835.6-183243 & $2.67^{+0.29}_{-0.37}$ & $2.80^{+0.34}_{-0.38}$ & $2.29^{+0.47}_{-0.38}$ & $2.67^{+0.47}_{-0.76}$ &  & $0.73^{+0.02}_{-0.02}$ & $0.72^{+0.02}_{-0.02}$ & $0.72^{+0.02}_{-0.02}$ \\
UCAC4 460-008869 & $3.25^{+0.20}_{-0.21}$ & $3.56^{+0.44}_{-0.36}$ & $3.09^{+0.20}_{-0.20}$ & $3.25^{+0.75}_{-0.36}$ &  & $0.80^{+0.02}_{-0.02}$ & $0.79^{+0.02}_{-0.02}$ & $0.80^{+0.02}_{-0.02}$ \\

\\
\hline
\end{tabular}

\end{table}

\section{Figures}

\begin{figure}
    \centering
    \includegraphics[width=\textwidth]{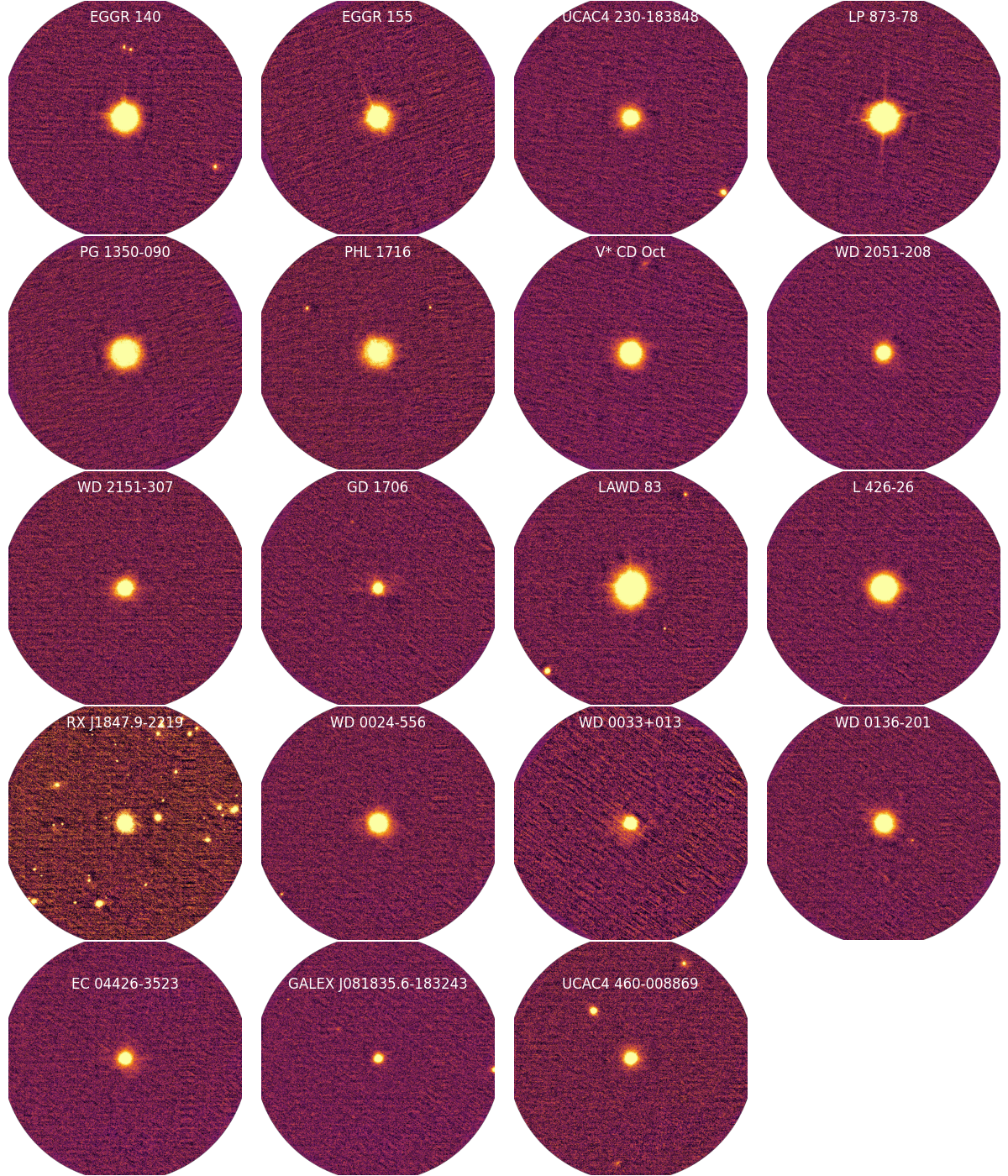}
    \caption{H-band images of the 19 WDs observed with the NIX camera of VLT/ERIS. All images are displayed with the same scale and orientation (east to the left, north up) as Fig. \ref{fig:followup}.}
    \label{fig: gallery}
\end{figure}

\end{appendix}

\end{document}